\documentclass[12pt]{article}
\usepackage{amsthm,amsmath,amssymb,bbm,bm}
\usepackage{mathtools}
\usepackage{natbib}
\usepackage[pdftex]{graphicx}
\usepackage{appendix}
\usepackage{caption}
\usepackage{subcaption}
\usepackage{url}
\usepackage[noend]{algpseudocode}
\usepackage{booktabs}
\usepackage{float}
\usepackage[usenames,dvipsnames,svgnames,table]{xcolor}
\usepackage[colorlinks=true,
            linkcolor=red,
            anchorcolor=blue,
            citecolor=blue,
            urlcolor=blue,
            hypertexnames=false]{hyperref}

\usepackage{setspace}
\usepackage{chngcntr}
\def\spacingset#1{\renewcommand{\baselinestretch}%
{#1}\small\normalsize} \spacingset{1}

\usepackage[a4paper, margin=1in]{geometry}

\usepackage{tikz}
\usetikzlibrary{bayesnet}
\usepackage{tikzscale}

\usepackage[labelformat=simple]{subcaption}

\counterwithin{equation}{section}
\theoremstyle{definition}

\counterwithin{thm}{section}

\counterwithin{rmk}{section}

\allowdisplaybreaks
\def\bal#1\eal{\begin{align}#1\end{align}}
\def\balnn#1\ealnn{\begin{align*}#1\end{align*}}

\floatstyle{boxed}
\newfloat{sampler}{thp}{los}
\floatname{sampler}{Algorithm}

\DeclareMathOperator*{\argmin}{arg\,min}

\DeclarePairedDelimiter\parentheses{\lparen}{\rparen}

\DeclarePairedDelimiter\set{\{}{\}}
\DeclarePairedDelimiterX{\norm}[1]{\lVert}{\rVert}{#1}
\DeclarePairedDelimiterX{\abs}[1]{\lvert}{\rvert}{#1}

\DeclarePairedDelimiterX{\expectarg}[1]{[}{]}{%
    \ifnum\currentgrouptype=16 \else\begingroup\fi
    \activatebar#1
    \ifnum\currentgrouptype=16 \else\endgroup\fi
}

\DeclarePairedDelimiterX{\variancearg}[1]{(}{)}{%
    \ifnum\currentgrouptype=16 \else\begingroup\fi
    \activatebar#1
    \ifnum\currentgrouptype=16 \else\endgroup\fi
}

\newcommand{\innermid}{\nonscript\;\delimsize\vert\nonscript\;}
\newcommand{\activatebar}{%
    \begingroup\lccode`\~=`\|
    \lowercase{\endgroup\let~}\innermid
    \mathcode`|=\string"8000
}

\newcommand{\Prob}{\mathbb{P} \, \variancearg}
\newcommand{\E}{\mathbb{E} \, \expectarg}

\newcommand{\condit}{ \mid }
\newcommand{\iidsim}{\overset{\text{iid}}\sim}

\newcommand{\Bern}{\operatorname{Bernoulli}}
\newcommand{\Bin}{\operatorname{Binomial}}
\newcommand{\Beta}{\operatorname{Beta}}
\newcommand{\Dir}{\operatorname{Dirichlet}}

\newcommand{\InvGamma}{\operatorname{\Gamma^{-1}}}

\newcommand{\GEM}{\operatorname{GEM}}
\newcommand{\DP}{\operatorname{DP}}

\def\mOne{{\mathbbm{1}}}
\newcommand{\ind}[1]{\mOne_{\{#1\}}}

\newcommand{\Reals}[1]{\mathbb{R}^{#1}}

\newcommand{\barm}{\bar{m}}
\newcommand{\barsigma}{\bar{\sigma}}

\newcommand{\X}{\mathbf{X}}

\newcommand{\bpi}{\boldsymbol\pi}
\newcommand{\bbeta}{\boldsymbol\beta}
\newcommand{\btheta}{\boldsymbol\theta}

\newcommand{\bdelta}{\boldsymbol\delta}

\newcommand{\bmu}{\boldsymbol\mu}
\newcommand{\bpsi}{\boldsymbol\psi}
\newcommand{\boldm}{\mathbf{m}}

\newcommand{\boldw}{\mathbf{w}}
\newcommand{\boldmbar}{\mathbf{\bar{m}}}
\newcommand{\boldmubar}{\bar{\bmu}}

\newcommand{\dyads}{\mathcal{D}}

\newcommand{\latentpos}{\mathcal{X}_{1:T}}

\newcommand{\labels}{\mathcal{Z}_{1:T}}

\newcommand{\latents}{\latentpos, \labels}

\newcommand{\clusterparams}{\bmu_{1:L}, \sigma_{1:L}^2}
\newcommand{\logit}{\operatorname{logit}}

\newcommand{\blind}{0}

\begin{document}

\if0\blind
{
\title{ A Bayesian Nonparametric Latent Space Approach to Modeling Evolving Communities in Dynamic Networks }

\author{
Joshua Daniel Loyal \thanks{Joshua Daniel Loyal is Ph.D Candidate (E-mail: \href{mailto:jloyal2@illinois.edu}{jloyal2@illinois.edu}), and Yuguo Chen is Professor (E-mail: \href{mailto:yuguo@illinois.edu}{yuguo@illinois.edu}), Department of Statistics, University of Illinois at Urbana-Champaign, Champaign, IL 61820. This work was supported in part by a grant from Sandia National Laboratories.} \and
Yuguo Chen
}

\date{\today}
\maketitle
}\fi
\if1\blind
{
\title{ A Bayesian Nonparametric Latent Space Approach to Modeling Evolving Communities in Dynamic Networks }

\author{\thanks{\vspace{2em}}}

\date{\today}
\maketitle
}\fi

\begin{abstract}
The evolution of communities in dynamic (time-varying) network data is a prominent topic of interest. A popular approach to understanding these dynamic networks is to embed the dyadic relations into a latent metric space. While methods for clustering with this approach exist for dynamic networks, they all assume a static community structure. This paper presents a Bayesian nonparametric model for dynamic networks that can model networks with evolving community structures. Our model extends existing latent space approaches by explicitly modeling the additions, deletions, splits, and mergers of groups with a hierarchical Dirichlet process hidden Markov model. Our proposed approach, the hierarchical Dirichlet process latent position clustering model (HDP-LPCM), incorporates transitivity, models both individual and group level aspects of the data, and avoids the computationally expensive selection of the number of groups required by most popular methods. We provide a Markov chain Monte Carlo estimation algorithm and apply our method to synthetic and real-world networks to demonstrate its performance.
\end{abstract}


\spacingset{1.2}

\section{Introduction} \label{sec:intro}


Many naturally occurring networks contain discrete changes in community structure. When high school students move across the country for college, old friendship groups often dissolve, leading the way for new friendship groups to form. After World War II, the Eastern and Western Blocs emerged and dominated the network of global alliances. However, after the fall of the Soviet Union, these blocs reshuffled into new political alliances. When exposed to external stimuli, regions of the brain activate before becoming dormant. By identifying these community-level phase changes, we can gain valuable insight into the rich processes that generate dynamic (longitudinal or time-varying) networks.



\begin{figure}[htb]
    \centering
    \begin{subfigure}[b]{0.32\textwidth}
    \includegraphics[width=\textwidth]{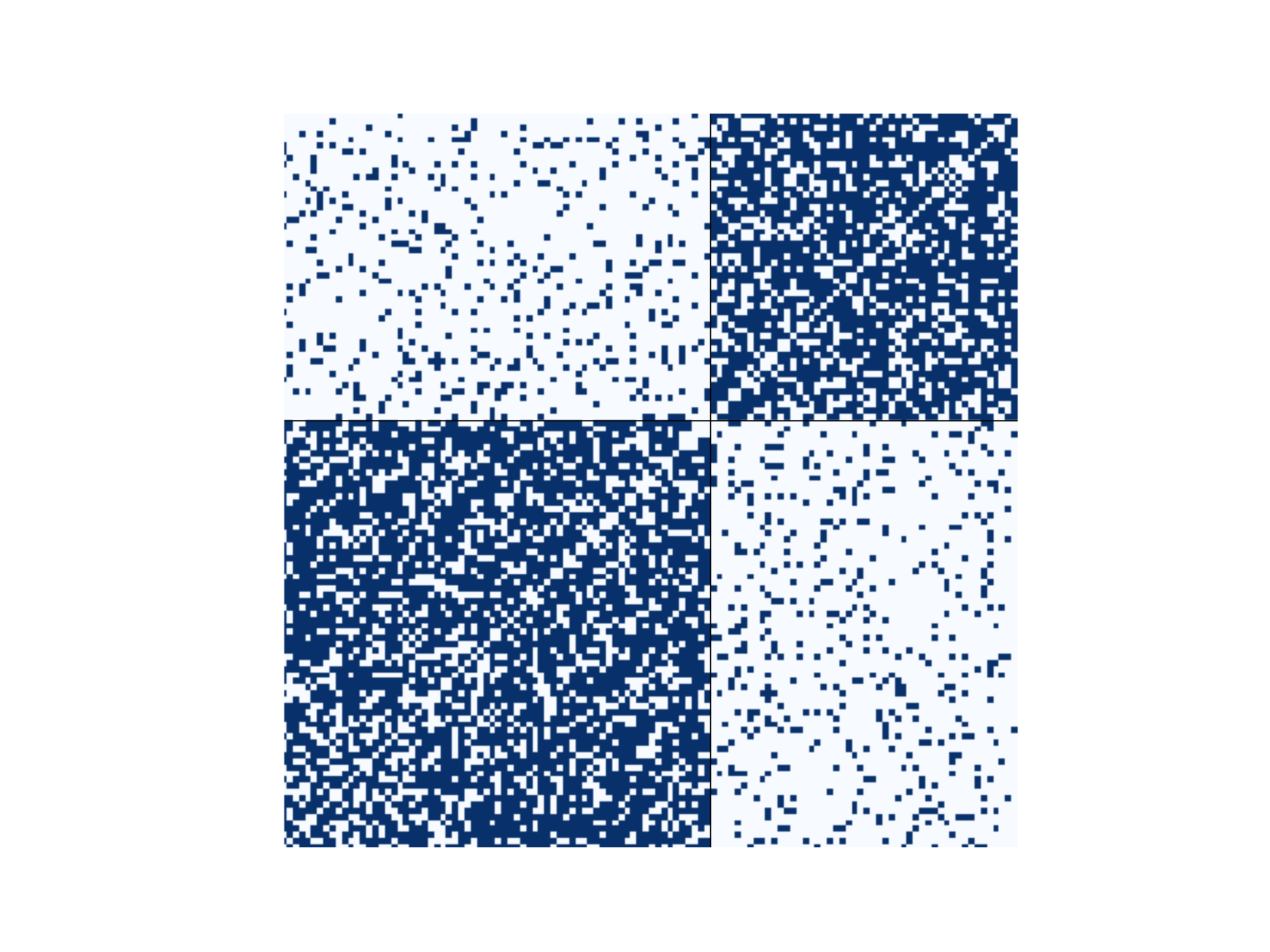}
    \caption*{$t = 1 - 3$}
    \end{subfigure}
    \hfill
    \begin{subfigure}[b]{0.32\textwidth}
    \includegraphics[width=\textwidth]{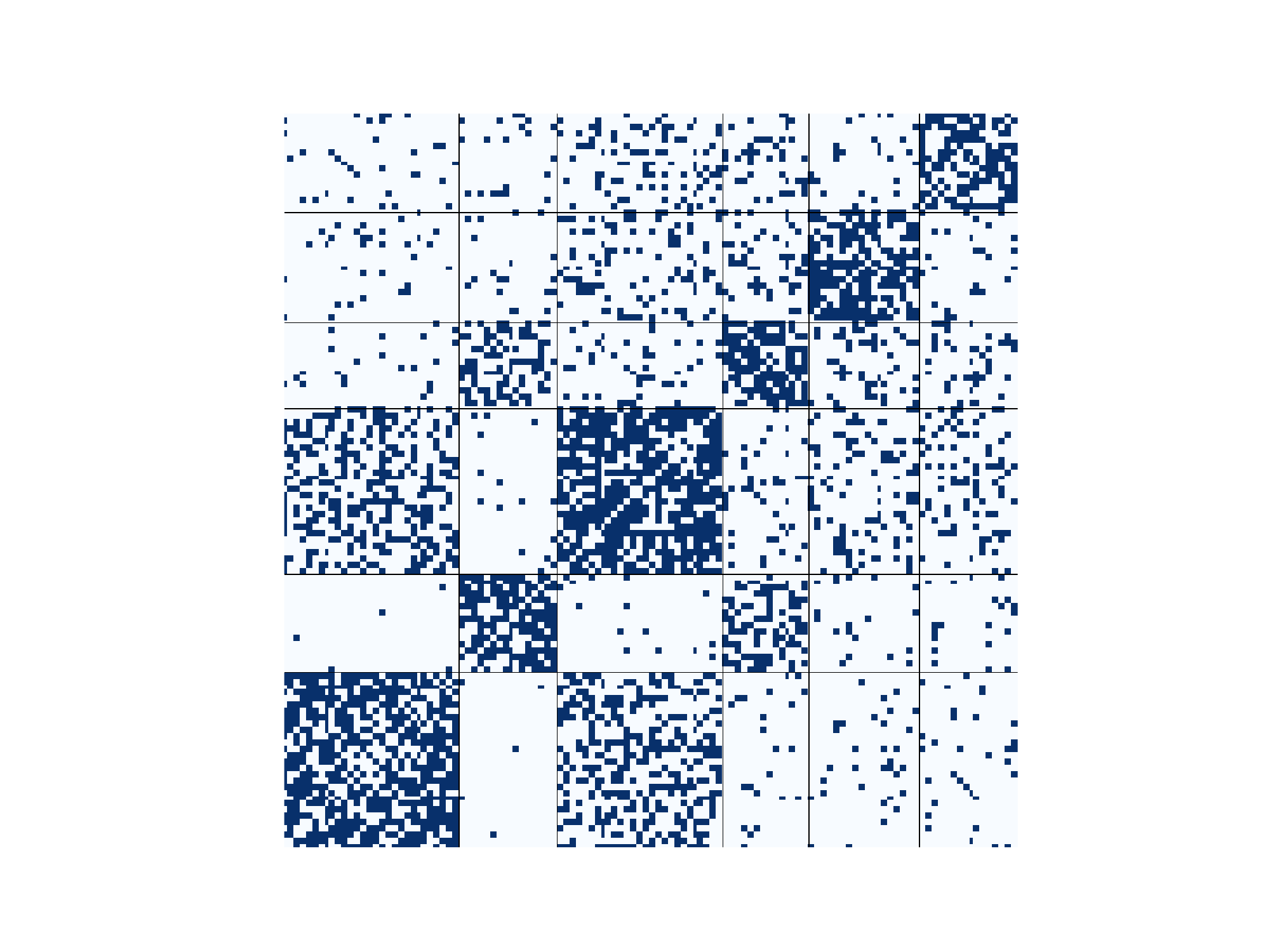}
    \caption*{$t = 4 - 6$}
    \end{subfigure}
    \hfill
    \begin{subfigure}[b]{0.32\textwidth}
    \includegraphics[width=\textwidth]{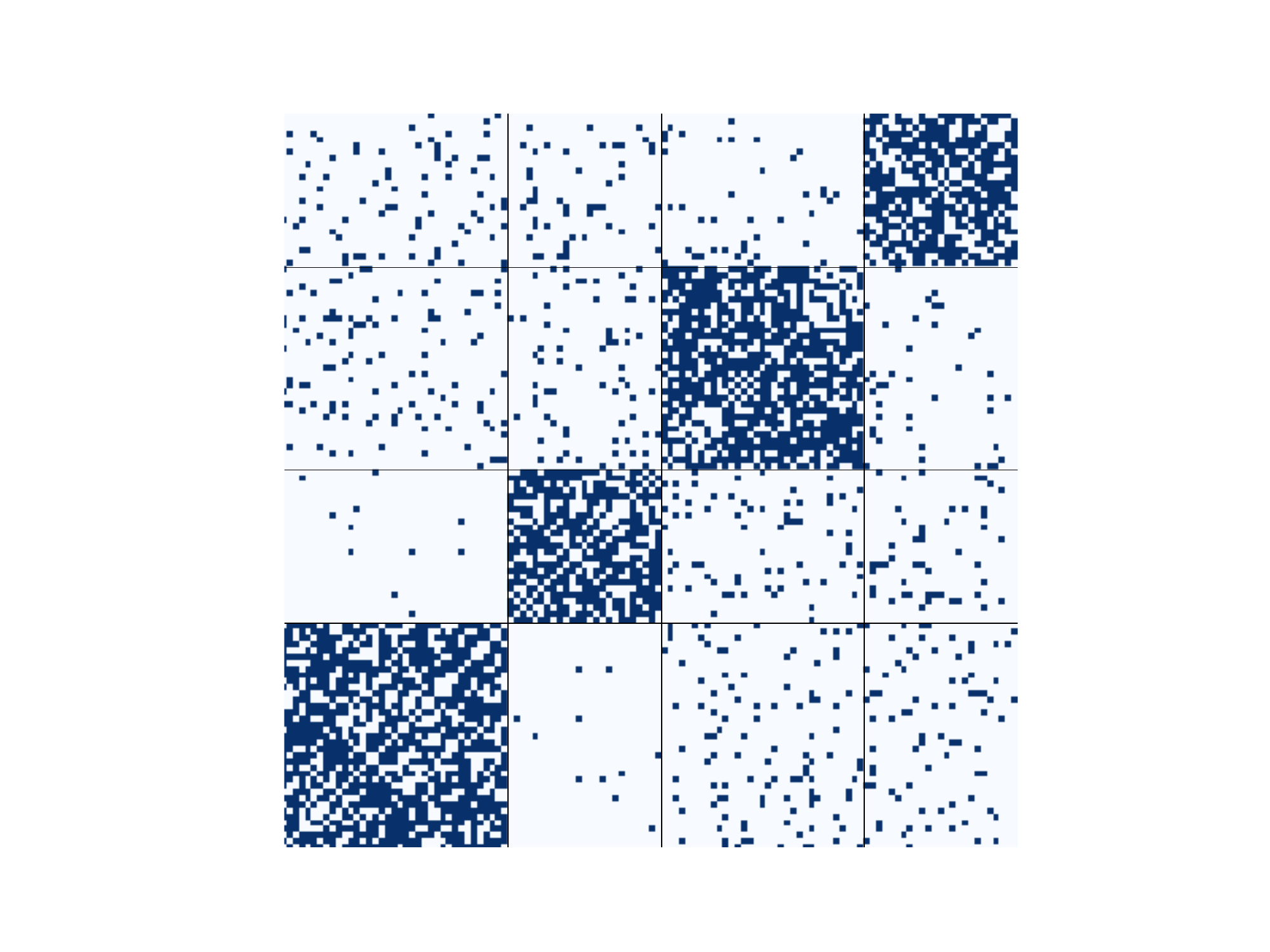}
    \caption*{$t = 7 - 9$}
    \end{subfigure}
    \caption{Adjacency matrices from a dynamic sequence of networks that exhibit an evolving community structure. At $t = 4$ the groups split from two into six groups, then at $t = 7$ the groups merge into four communities. The rows and columns of each adjacency matrix have been sorted according to the ground truth labels. These networks were simulated according to the procedure described in Section \ref{subsec:inhomogeneous_sim}.}
    \label{fig:evolving_communities}
\end{figure}

In this work, we address this problem of inferring discrete changes in a network's community structure. See Figure \ref{fig:evolving_communities} for a concrete example.
In this case, a dynamic network's community structure contains two groups during the first three time points, splits into six groups at the fourth time point, and then merges into four groups during the seventh time point. These additions, deletions, splits, and mergers of groups in dynamic networks is our primary object of interest. For brevity, we refer to these group-level dynamics as a network's \textit{evolving community structure}. Note that this does not refer to individual actors (or nodes) moving between existing communities. While we will model such actor-level dynamics, our primary goal is to infer changes in community structure that occur on a larger macro scale.

To infer an evolving community structure in dynamic networks, we adopt the latent space approach to network modeling. Originally developed in \citet{hoff2002}, latent space models (LSMs) embed actors within a Euclidean space (the distance model) or a hypersphere (the projection model). Closeness in the latent space increases the probability that the two actors form an edge in the observed network. For this reason, one interprets the proximity of two actors in the latent space as an indication that they have similar characteristics. The LSM's popularity stems from the intuitive meaning of the embeddings and its ability to naturally incorporate desirable sociological features such as homophily, reciprocity, and transitivity.  The distance model was extended to undirected dynamic networks by \citet{sakar2006} and directed dynamics networks by \citet{sewell2015}.

The latent space model was first applied to the community detection problem in the case of a single static network by \citet{handcock2007}. Their proposed latent position clustering model (LPCM) uses a Gaussian mixture model to cluster the latent positions embedded according to the Euclidean distance model. The idea is that a cluster of actors in the latent space will correspond to a densely connected community in the network. This work was extended to dynamic networks by \citet{sewell2017}. Their approach allows the actors to move between a static number of groups at each point in time. They accomplish this by endowing the actor's latent trajectories with autoregressive hidden Markov model (AR-HMM) dynamics.

The LPCM is an excellent model for community detection in dynamic networks, but it has two key shortcomings that make it inadequate for inferring an evolving community structure: (1) the number of communities must be set a priori so that uncertainty in the community structure cannot be assessed in a fully Bayesian way, and (2) the number of communities must remain static over time. The first shortcoming is important since multiple clusterings often fit the data well, which means we need a way to compare uncertainty across these partitions. The second shortcoming implies that existing LPCMs are unable to allow the number of communities to change over time so that inference over evolving communities is not possible. The current inability of LPCMs to handle time-varying groups in longitudinal networks motivates our investigation into a suitable Bayesian methodology.

In this work, we propose a Bayesian nonparametric model that addresses the shortcomings outlined in (1) and (2). As in \citet{sewell2017}, we model the actor's trajectories through the latent space as a collection of independent and identically distributed (iid) AR-HMMs with Gaussian emissions. However, unlike their approach, we place a hierarchical Dirichlet process (HDP) prior on the Markov chain's transition distributions to allow for inference over the number of communities. Furthermore, we no longer assume that the hidden label Markov chain is homogeneous. Instead, we allow for time-inhomogeneous transition distributions. This modification provides us with the flexibility to model the additions, deletions, splits, and mergers of communities.
Since our approach combines the advantages of the HDP with the LPCM for modeling dynamic networks, we refer to our model as the hierarchical Dirichlet process latent position clustering model (HDP-LPCM). To the best of our knowledge, there is no other latent space approach that accomplishes our modeling criteria.

The rest of this paper is organized as follows. Section \ref{sec:rel_methods} reviews other Bayesian nonparametric approaches for modeling evolving communities. In Section \ref{sec:models}, we describe the hierarchical Dirichlet process latent position clustering model (HDP-LPCM), our proposed Bayesian nonparametric method for modeling evolving communities in dynamic networks. In Section \ref{sec:estimation}, we elaborate on the details of our Metropolis-Hastings within Gibbs estimation procedure as well as our methodology for posterior summarization. We demonstrate the empirical performance of our proposed method through various simulation studies in Section \ref{sec:simulation}. In Section \ref{sec:real_data}, we use our model to analyze two real-world dynamic networks: a network of military alliances during the Cold War \citep{gibler2009} and a narrative network constructed from the Game of Thrones television series \citep{beveridge2018}. Section \ref{sec:disc} contains a discussion.

\section{Related Methods}\label{sec:rel_methods}
Although modeling evolving communities in dynamic networks is new to the latent space methodology, the statistical modeling of varying communities has a long history in the broader network literature. For brevity, we focus on Bayesian nonparametric approaches because they are most related to our work. We restrict the discussion to the three most popular varieties of statistical network models: stochastic block models, mixed-membership stochastic block models, and latent feature models.

For stochastic block models, \citet{kemp2006} introduced the infinite relational model (IRM), which uses a Dirichlet process prior to infer the number of blocks in the traditional stochastic block model. \citet{ishiguro2010} extended this work to longitudinal networks in their dynamic infinite relational model (dIRM). Similar to our work, the dIRM assumes the block memberships form a time-inhomogeneous Markov chain. In the mixed-membership literature, \citet{fan2015} introduced the dynamic infinite mixed-membership stochastic block model. Their model also utilizes the hierarchical Dirichlet process framework to extend the original mixed-membership model \citep{airoldi2008mixed, fu2009, ho2011} to a dynamic setting. In contrast to our method, they do not assume an HMM structure and only use the HDP to re-sample the mixed-membership vectors at every time step. Finally, in the latent feature modeling literature, \citet{kim2013} introduced the nonparametric multi-group membership model. They utilize a distance-dependent Indian buffet process (dd-IBP) to assign each actor a latent binary feature vector. Due to their nonparametric prior, they can infer an evolving number of binary features.

While some of these competing approaches have addressed shortcomings (1) and (2), LPCMs have many appealing advantages that make our extension worthwhile. Foremost is the LPCM's ability to capture network structures on multiple scales. Unlike the block models described above, which assume connections are purely governed by a global clustering, latent space models allow for heterogeneous connectivity patterns due to an actor's local position. Furthermore, unlike other methods, the latent space provides an interpretable visualization of the entire network. In practice, these visualizations are crucial for allowing domain experts to make qualitative assessments and critiques of the statistical methodology.

\section{The HDP Latent Position Clustering Model} \label{sec:models}

We consider binary relational data between $n$ individuals recorded over $T$ time periods. These relations are collected in a sequence of $n \times n$ binary adjacency matrix $Y_t$, $t = 1, \dots, T$, where the entries $y_{ijt}$ indicate the presence ($y_{ijt} = 1$) or absence ($y_{ijt} = 0$) of an edge between individuals $i$ and $j$ at time $t$. For clarity, we only consider undirected random graphs without self-loops, so that $Y_t$ is a symmetric matrix with zeros on the diagonal. The extension of our model to directed networks or networks with weighted edges is straightforward and discussed in Section \ref{sec:disc}.


To each of the $n$ individuals, we associate a latent (unobserved) position that may vary through time in a $p$-dimensional Euclidean latent space. We represent the latent position of individual $i$ at time $t$  with the vector-valued random variable $\X_t^i \in \Reals{p}$. In addition, we collect a snapshot of all individual latent positions at time $t$ in the $n \times p$ matrix $\mathcal{X}_t = (\X_t^1, \dots, \X_t^n)'$. Like the traditional LPCM, we assign each actor to a latent group at each time point. Note that their assignment may change over time. However, unlike the LPCM, we assume that the number of groups changes over time to accommodate the network's evolving community structure. We use $G$ to denote the total number of groups in the network over all observational periods. We refer to a group that contains at least one actor at time $t$ as an \textit{active group at time $t$}. We use $G_t \subseteq \set{1, \dots, G}$ to indicate the set of all active groups at time $t$. Note that under the assumption of an evolving community structure, $G_t$ is a random set and may grow or shrink over time. We encode the latent group membership of individual $i$ at time $t$ with the integer-valued random variable $Z_t^i \in G_t$. The collection of group assignments for all individuals at time $t$ is summarized by the $n$-dimensional vector $\mathcal{Z}_t = (Z_t^1, \dots, Z_t^n)^{\prime}$.

Common to the latent space literature, we assume that the latent community labels only play a role in the distribution of the latent positions.
Formally, we assume that the formation of edges in the dynamic networks $Y_{1:T}$ is conditionally independent of the actor labels $\mathcal{Z}_{1:T}$ given the latent positions $\mathcal{X}_{1:T}$, i.e., $Y_{1:T} \perp \mathcal{Z}_{1:T} \condit \mathcal{X}_{1:T}$. Note that throughout the remainder of this work, we will use the notation $A_{1:K}$ to refer to the sequence $(A_1, A_2, \dots, A_K)$. This allows us to decompose the joint probability distribution as follows:
\begin{equation}\label{eq:prob_factorization}
\Prob{Y_{1:T}, \mathcal{X}_{1:T}, \mathcal{Z}_{1:T}} = \Prob{Y_{1:T} | \latentpos} \, \Prob{\latentpos, \labels}.
\end{equation}
This independence assumption says that the probability of a tie is solely determined by the underlying latent positions of the actors. This decomposition is consistent with the latent space clustering idea described earlier. Specifically, the notion that an underlying cluster of actors in the latent space results in observed communities. We believe such a generative model is natural for modeling communities in networks. For example, friend groups in social networks often form due to the similar interests or characteristics of their members.



\subsection{The Euclidean Distance Model}\label{subsec:distance_models}
As in \citet{hoff2002}, we posit that the probability of an edge forming between actors increases as the Euclidean distance between actors decreases. Let $d_{ijt} = \norm{\X_{t}^i - \X_{t}^j}_2$ denote the Euclidean distance between actors $i$ and $j$ at time $t$. A conventional link between the conditional probability of forming an edge and $d_{ijt}$ is the logistic regression model:
\begin{equation}\label{eq:adjacency_model}
\Prob{Y_{ijt} = 1 | \mathcal{X}_t, \bpsi} = \logit^{-1}(\eta_{ijt}) = \frac{\exp(\eta_{ijt})}{1 + \exp(\eta_{ijt})},
\end{equation}
where $\eta_{ijt}$ is a linear predictor that depends on the distances $d_{ijt}$ and the vector $\bpsi$ holds any additional parameters. A sequence of conditional independence assumptions results in the full network likelihood. First, we assume that the longitudinal networks are conditionally independent given the latent positions, i.e.,  $Y_t \perp Y_s \condit \mathcal{X}_{1:T}$ for all $s, t \in \set{1, \dots, T}$. Second, we posit that edges between actors form independently conditioned on their latent positions at each time point.
Under these assumptions, the likelihood of the adjacency matrices factors as a product over the networks at each time point and the set of dyads $\dyads$:
\begin{equation}\label{eq:network-likelihood}
\Prob{Y_{1:T} | \mathcal{X}_{1:T}, \bpsi} = \prod_{t=1}^T \prod_{(i, j) \in \dyads} \Prob{Y_{ijt} = y_{ijt} | \mathcal{X}_t, \bpsi} =\prod_{t=1}^T \prod_{(i,j) \in \dyads} \frac{\exp(y_{ijt}\eta_{ijt})}{1 + \exp(\eta_{ijt})}.
\end{equation}
In this work, we focus on undirected networks without self-loops for which
\begin{equation}
\dyads = \set{(i, j) : 1 \leq i \leq n, j < i}
\end{equation}
and the linear predictor takes the form
\begin{equation}\label{eq:undirected_predictor}
\eta_{ijt} = \beta_0 - d_{ijt}
\end{equation}
so that $\bpsi = \set{\beta_0}$. Various versions of this likelihood have been proposed, such as in \citet{sakar2006}, \citet{krivitsky2009}, and \citet{sewell2015}. The intercept parameter $\beta_0$ is sometimes referred to as the abundance. This is because higher values of $\beta_0$ result in a higher probability of forming edges.


\subsection{An AR-HMM for Latent Space Dynamics}\label{subsec:hmm-network}

Recall that our goal is to endow the dynamics of the latent space with a probabilistic structure that allows for inference over an evolving collection of communities. To accomplish this goal, we must place an adequately flexible joint distribution over the latent space, $\Prob{\mathcal{X}_{1:T}, \mathcal{Z}_{1:T}}$. To begin, we adopt the assumptions taken by \citet{sewell2017}. Specifically, we assume that an actor's community assignments form a Markov chain, i.e.,
\begin{equation}\label{eq:label_markov}
Z_t^i \condit Z_1^i, \dots, Z_{t-1}^i \overset{\mathcal{D}}= Z_{t}^i \condit Z_{t-1}^i,
\end{equation}
where $\overset{\mathcal{D}}=$ denotes equality in distribution. Similarly, we assume an actor's latent position follows a Markov process with an additional dependence on the current community assignment, i.e.,
\begin{equation}\label{eq:pos_markov}
\X_{t}^i \condit \X_{1}^i, \dots, \X_{t-1}^i, Z_{1}^i, \dots, Z_{t-1}^i, Z_{t}^i \overset{\mathcal{D}}= \X_{t}^i \condit \X_{t-1}^i, Z_{t}^i.
\end{equation}
These two assumptions allow us to factor the marginal density of an individual actor's trajectory. In particular, we conclude that an actor's trajectory follows an independent autoregressive hidden Markov model (AR-HMM):
\begin{equation}\label{eq:morkov_model}
p(\X_{1:T}^i,Z_{1:T}^i) = p(Z_1^i)p(\X_1^i \condit Z_1^i) \prod_{t=2}^T p(Z_t^i \condit Z_{t-1}^i) p(\X_t^i \condit \X_{t-1}^i, Z_t^i),
\end{equation}
where $p(Z_1^i)$ is the initial distribution over actor $i$'s initial community assignment, $p(Z_t^i \condit Z_{t-1}^i)$ is the transition distribution between actor $i$'s label assignment at time $t-1$ and time $t$, and $p(\X_t^i \condit \X_{t-1}^i, Z_t^i)$ is the emission distribution for actor $i$'s latent position at time $t$, which we allow to depend on the previously emitted value $\X_{t-1}^i$.

To properly model evolving communities, we make an important departure from previous dynamic latent space models. In particular, we expand the probabilistic model to include time-inhomogeneous Markov chains where the transition matrix $p(Z_t^i \condit Z_{t-1}^i)$ can vary over time. This is in contrast with previous methods, which assume that the probability for an actor to transition from community $i$ to community $j$ is the same at all times. We argue that time-inhomogeneous transitions are essential characteristics of evolving communities. For example, the addition of a group requires the transition matrices to add a non-zero probability to transition into that new group. Furthermore, once the new group is added to the network, an actor's probability to transition into that group may approach a different steady-state than when it initially appeared. Although models that utilize homogeneous Markov chains may infer changing groups through the inclusion of empty clusters, their inferences will certainly be biased due to smoothing over sudden changes in group structure.


Tractable inference over these time-inhomogeneous latent trajectories is made possible by tying together the transition distributions and emission distributions of every actor.
We collect the transition probabilities into a single row-stochastic transition matrix $\Pi_t$ for each time step $t = 2, \dots, T$. Each entry is defined as $(\Pi_t)_{jk} = \pi^{t}_{jk} = p(Z_t^i = k \condit Z_{t-1}^i = j)$, which is the same for all $n$ actors. In what follows, it will be useful to isolate the $j$-th row of $\Pi_t$ in the vector $\bpi_j^t$. Furthermore, we refer to $\set{\bpi_j^t}$ as the set of rows of the transition matrix at time $t$.
For simplicity, we assume that all actors share a common initial state $Z_0^i = 0$. This allows us to view the initial distribution of the Markov chain as a special transition distribution denoted by $\bpi_0^1 = (\pi_{01}^1, \dots, \pi_{0 \, G_0}^1)$, where $\pi_{0k}^1 = p(Z_1^i = k \condit Z_0^i = 0) = p(Z_1^i = k)$ is equal for all actors.



Conditioned on the initial and transition probability matrices as well as the parameters of the emission distributions $\btheta$, the data generating process of all the actor trajectories is characterized by a collection of iid AR-HMMs. In this work, we model the joint density over all actor trajectories as
\begin{equation}\label{eq:latent-community-model}
\begin{split}
p(\mathcal{X}_{1:T}, \mathcal{Z}_{1:T} \condit \Pi_{2:T}, \bpi_0^1, \btheta) &= \prod_{i=1}^n p(\X_{1:T}^i, Z_{1:T}^i \condit \Pi_{2:T}, \bpi_0^1, \btheta) \\
                                       &= \prod_{i=1}^n \pi_{0,Z_{1}^i}^1 \, N(\X_{1}^i \condit \bmu_{Z_1^i}, \sigma^2_{Z_1^i} I_p) \\
                                       &\qquad \times \prod_{t=2}^T \pi^t_{Z_{t-1}^i, Z_t^i} \, N(\X_t^i \condit \lambda \bmu_{Z_t^i} + (1 - \lambda) \X_{t-1}^i, \sigma^2_{Z_t^i} I_p),
\end{split}
\end{equation}
where $I_p$ is the $p \times p$ identity matrix, and $N(\X \condit \bmu, \Sigma)$ is the normal density with mean vector $\bmu$ and covariance $\Sigma$ evaluated at the point $\X$. Note that in this case $\btheta = \set{\set{\bmu_g, \sigma_g^2}_{g=1}^G, \lambda}$. Like the clustering model in \citet{handcock2007}, the communities are modeled as a multivariate normal distribution in the latent space with mean location $\bmu_g$, and spherical covariance $\sigma^2_g \, I_p$. As in the longitudinal clustering approach of \citet{sewell2016}, the mean position $\X_t^i$ is equal to $\lambda \bmu_g + (1 - \lambda)\X_{t-1}^i$ where $\lambda \in (0, 1)$. This is a blend between the actor's previous position and the current assigned group location. Consequently, $\lambda$ is called the blending coefficient.



To complete the model, we must specify a prior on the initial distribution $\bpi_0^1$, the transition probabilities $\Pi_{2:T}$, and the parameters of the emission distribution $\btheta = \set{\set{\bmu_g, \sigma_g^2}_{g=1}^G, \lambda}$.
A parametric approach is taken in \citet{sewell2017}. Their model restricts itself to time-homogeneous Markov chains and assumes a fixed number of communities throughout time. As previously explained, their model is inadequate for inference over an evolving community structure. They are neither able to quantify uncertainty in the number of groups nor account for time-inhomogeneous transitions. For this reason, we turn to the flexibility afforded by a Bayesian nonparametric approach.


%
%
%


\subsection{Background: Dirichlet Processes and the HDP-HMM}


%
%

The Dirichlet process (DP) is a distribution over discrete probability measures:
\begin{equation}
\bbeta \sim \GEM(\gamma), \qquad \btheta_k \iidsim H, \qquad G_0 = \sum_{k=1}^{\infty} \beta_k \delta_{\btheta_k},
\end{equation}
where $\bbeta \sim \GEM(\gamma)$ denotes the sticking-breaking process \citep{sethuraman1994}:
\begin{equation}
\beta_k = w_k \prod_{i=1}^{k-1}(1 - w_i), \qquad w_k \sim \Beta(1, \gamma), \qquad k = 1, 2, \dots \, .
\end{equation}
From now on we will use the notation $G_0 \sim \DP(\gamma, H)$ to indicate draws from a DP with concentration parameter $\gamma$ and base measure $H$. Since draws from a DP are discrete with probability one, the DP cannot be used as a general nonparametric prior over continuous densities. To extend the DP to continuous density estimation, one often uses $G_0$ as a mixing measure over some parametric class of distributions $f_{\btheta}$. This construction is known as the DP mixture model. The sampling process  begins by drawing indicator variables $z_i \iidsim \bbeta$, and generating observations as $y_i \sim f_{\btheta_{z_i}}$. Note that in this work, we adopt the convention that when $\bbeta$ represents a distribution over the natural numbers, i.e., $\sum_{j=1}^K \beta_j = 1$, then we use $z \sim \bbeta$ to mean $z \sim \sum_{j=1}^K \beta_j \, \delta(z, j)$, where $\delta(i,j)$ is the Kronecker delta.


The hierarchical Dirichlet process (HDP) is a distribution over a collection of discrete probability measures that share a common set of atoms:
\begin{equation}\label{eq:hdp_model}
\begin{split}
\bbeta &\sim \GEM(\gamma), \\
\bpi_j &\iidsim \DP(\alpha, \bbeta), \qquad \btheta_k \iidsim H, \qquad G_j = \sum_{k=1}^{\infty} \pi_{jk} \delta_{\btheta_k}, \qquad j = 1, \dots, J.
\end{split}
\end{equation}
The group-specific distributions of the HDP, $\bpi_j$, are often used as priors over the rows of an infinite dimensional transition matrix, i.e., $(\Pi)_{jk} = \pi_{jk}$ for $j \in \mathbb{N}$. This formulation allows for the construction of the hierarchical Dirichlet process hidden Markov model (HDP-HMM) \citep{teh2006}, which is a natural Bayesian nonparametric extension of the Bayesian HMM \citep{robert2000}. The sampling mechanism of the HDP-HMM proceeds as follows: one samples the hidden states sequentially as $z_t \sim \bpi_{z_{t-1}}$, and the observations are linked to the global parameters via $y_t \sim f_{\btheta_{z_t}}$. Note that the sharing of atoms induced by the HDP prior allows the Markov chain to utilize a single global set of parameters, $\btheta_k$, at all time points.


Despite its popularity the original HDP-HMM struggles to model Markov chains with long state durations. To remedy this issue \citet{fox2011a} introduced the sticky hierarchical Dirichlet process (sticky HDP):
\begin{equation}
\begin{split}
\bbeta &\sim \GEM(\gamma), \\
\bpi_j & \iidsim \DP(\alpha + \kappa, \frac{\alpha \bbeta + \kappa \bdelta_j}{\alpha + \kappa}),  \qquad  \btheta_k \iidsim H, \qquad G_j = \sum_{k=1}^{\infty} \pi_{jk} \delta_{\btheta_k}, \qquad j = 1, \dots, J \, ,
\end{split}
\end{equation}
where $\bdelta_j$ is a vector of zeros except for a single one at the $j$-th index. Analogous to the HDP, the sticky HDP can be used as a prior over the transition matrices of an HMM. In this case, the extra stickiness parameter $\kappa > 0$ biases the process towards self-transitions. As a result, \citet{fox2011a} found that the corresponding sticky HDP-HMM better models the longer state durations found in real world applications. For this reason, we chose to use the sticky variant of the HDP as a prior over the transition matrices in Equation (\ref{eq:latent-community-model}).

\subsection{An HDP Prior for Time-Inhomogeneous Markov Chains}\label{subsec:time-inhomo}

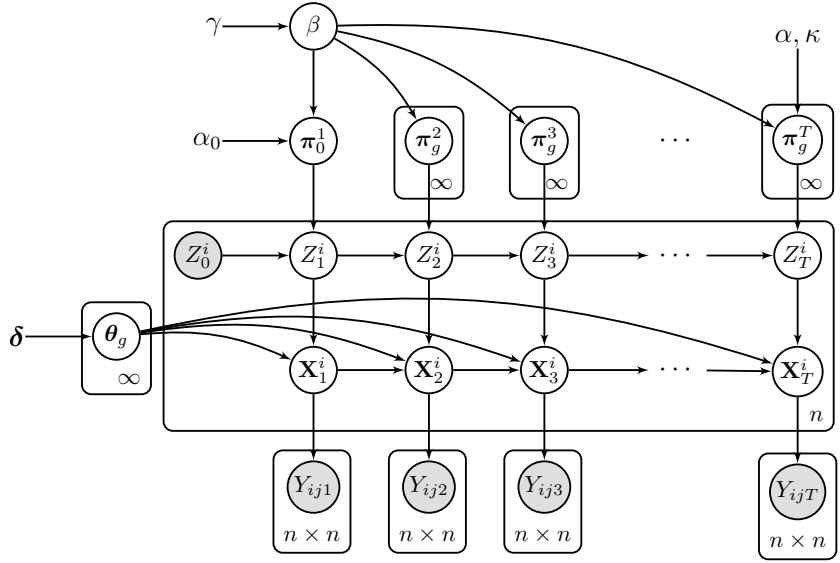
\begin{figure}[htbp]
\centering
\resizebox{0.7\textwidth}{!}{
\begin{tikzpicture}[node distance=2cm, auto,>=latex', thick, scale = 0.5]

\node[latent] (z1) {$Z_1^i$};
\node[obs, left=of z1] (z0) {$Z_0^i$};
\edge{z0} {z1};

\node[latent, above=of z1] (pi1) {$\bpi_{0}^1$};

\edge{pi1} {z1};

\node[latent, below=of z1] (x1) {$\X_1^i$};
\node[obs, below=of x1] (Y1) {$Y_{ij1}$};

\node[latent, below left=of z0] (theta) {$\btheta_g$};

\plate {theta_plate} {(theta)} {$\infty$};
\node[const, left=of theta] (delta) {$\bdelta$};
\edge{delta} {theta};

\node[latent, right=of z1] (z2) {$Z_2^i$};
\node[latent, below=of z2] (x2) {$\X_2^i$};
\node[obs, below=of x2] (Y2) {$Y_{ij2}$};

\node[latent, above=of z2] (pi2) {$\bpi^{2}_{g}$};
\edge{pi2} {z2};
\plate {pi2_plate} {(pi2)} {$\infty$};

\node[latent, right=of z2] (z3) {$Z_3^i$};
\node[latent, below=of z3] (x3) {$\X_3^i$};
\node[obs, below=of x3] (Y3) {$Y_{ij3}$};

\node[latent, above=of z3] (pi3) {$\bpi^{3}_{g}$};
\edge{pi3} {z3};
\plate {pi3_plate} {(pi3)} {$\infty$};

\node (dots1) [right of=z3] {$\cdots$};
\node (dots2) [right of=x3] {$\cdots$};
\node (dotspi) [right of=pi3] {$\cdots$};

\node[latent, right=of dots1] (zT) {$Z_T^i$};
\node[latent, below=of zT] (xT) {$\X_T^i$};
\node[obs, below=of xT] (YT) {$Y_{ijT}$};

\node[latent, above=of zT] (piT) {$\bpi^{T}_{g}$};
\edge{piT} {zT};
\plate {piT_plate} {(piT)} {$\infty$};

\node[latent, above=of pi1] (beta) {$\beta$};
\node[const, left=of beta] (gamma) {$\gamma$};
\edge {beta} {pi1};
\edge {gamma} {beta};

\edge{z1} {x1};
\edge{z1} {z2};
\edge{x1} {x2};
\edge{x1} {Y1};
\edge{x2} {Y2};
\edge{x3} {Y3};
\edge{xT} {YT};
\plate {plate1} { (Y1) } {$n \times n$};

\edge{z2} {x2};
\edge{z2} {z3};
\edge{x2} {x3};
\plate {plate2} { (Y2) } {$n \times n$};

\edge{z3} {x3};
\edge{z3} {dots1};
\edge{x3} {dots2};
\plate {plate3} { (Y3) } {$n \times n$};

\edge{dots1} {zT};
\edge{dots2} {xT};
\edge{zT} {xT};

\node[const, left=of pi1] (alpha0) {$\alpha_0$};
\node[const, above=of piT] (alpha) {$\alpha, \kappa$};
\edge {alpha0} {pi1};
\edge {alpha} {piT};

\plate {plate4} { (YT) } {$n \times n$};

\plate {platehmm} {(z0) (z1) (z2) (z3) (zT) (x1) (x2) (x3) (xT)} {$n$};

\path
    (theta) edge[->, bend left=15] node [right] {} (x1);
\path
    (theta) edge[->, bend left=15] node [right] {} (x2);
\path
    (theta) edge[->, bend left=15] node [right] {} (x3);
\path
    (theta) edge[->, bend left=15] node [right] {} (xT);

\path
    (beta) edge[->, bend left=15] node [right] {} (pi2);
\path
    (beta) edge[->, bend left=15] node [right] {} (pi3);
\path
    (beta) edge[->, bend left=15] node [right] {} (piT);

\end{tikzpicture}
}
\caption{The graphical model for the HDP-LPCM. The dependence on $\alpha$ and $\kappa$ is only displayed for the last transition matrix, although it is present in all transition matrices from $t = 2, \dots, T$. This change is for clarity only. The HDP-LPCM consists of a collection of actor specific iid time-inhomogeneous AR-HMMs whose shared state space is inferred by an HDP. The latent positions are linked to the observed network through the Euclidean distance model of \citet{hoff2002}.}
\label{fig:graphical-model}
\end{figure}

We now present our extension to the latent position clustering model that can infer an evolving community structure in dynamic networks. To accomplish this objective, we place the following Bayesian nonparametric prior on the transition matrices in Equation (\ref{eq:latent-community-model}):
\begin{equation}\label{eq:inhomo_prior}
\begin{split}
\bbeta &\sim \GEM(\gamma),  \\
\bpi_0^1 &\iidsim \DP(\alpha_0, \bbeta), \\
\bpi_g^{t} &\iidsim \DP(\alpha + \kappa, \frac{\alpha \bbeta + \kappa \bdelta_g}{\alpha + \kappa}), \qquad (\bmu_g, \sigma_g^2) \iidsim H \times F, \qquad g = 1, 2, \dots, \quad t = 2, \dots, T.
\end{split}
\end{equation}
In this work, we take $H$ to be a multivariate normal distribution and $F$ to be an inverse gamma distribution, i.e., $\bmu_g \iidsim N(\bmu_0, \tau^2 I_p)$ and $\sigma_g^2 \iidsim \InvGamma(a/2, b/2)$. For the initial distribution, we set $\kappa = 0$ and allocate it a separate concentration parameter $\alpha_0$ to distinguish it from the other transition distributions. Note that, unlike the traditional HDP-HMM,
our model re-samples the rows of the transition matrix, $\bpi_g^t$, at each time step. This difference allows us to infer time-inhomogeneous Markov chains. As we previously argued, this time-inhomogeneity is crucial for modeling evolving communities. In short, our model extends the dynamic Euclidean distance model proposed in \citet{sewell2017} by adding a nonparametric HDP prior over time-inhomogeneous transition matrices. For this reason, we refer to our model as the hierarchical Dirichlet process latent position clustering model (HDP-LPCM). A graphical representation of the HDP-LPCM is depicted in Figure \ref{fig:graphical-model}.

Although our prior borrows heavily from the sticky HDP-HMM formulation, there are three key differences between the state-space model utilized in the HDP-LPCM and the traditional sticky HDP-HMM. First, the emitted values $\X_t^i$ have an autoregressive dependence. This modification was originally explored outside the network setting by \citet{fox2011b} to model switching linear dynamical systems with Gaussian observations. However, in our case, the observations are a sequence of binary adjacency matrices. The second difference is that the HDP-HMM contains a single state sequence $Z_{1:T}$, while we infer multiple state sequences $Z_{1:T}^i$ for $i = 1, \dots, n$ from a single observed sequence of networks. The final departure from the HDP-HMM is the re-sampling of the rows of the transition matrix $\set{\bpi_g^t}$ at every time step. This difference introduces time-inhomogeneity into the Markov chain, which is crucial for modeling the additions, deletions, splits, and mergers of groups found in evolving communities.




In addition to properly modeling an evolving community structure, the other important accomplishment of our model is its ability to infer the number of communities from the data in a fully Bayesian way. A typical parametric approach (specifying $\abs{G_t}$ at each time step) would require comparing a combinatorial amount of models. This task is computationally infeasible for even a small number of groups and time steps. Through a nonparametric prior, we naturally incorporate model selection into our inference procedure, which avoids the computationally expensive model selection step found in many LPCMs \citep{handcock2007, sewell2017}.

\subsection{The Weak-Limit Approximation}
For inference, we utilize the weak-limit approximation to the HDP proposed in \citet{fox2011a}. The approximation replaces the infinite dimensional DPs with finite $L$-dimensional Dirichlet distributions as follows:
\begin{equation}\label{eq:weak_limit}
\begin{split}
\bbeta &\sim \Dir(\gamma/L, \dots, \gamma/L), \\
\bpi_0^{1} &\iidsim  \Dir(\alpha_0 \beta_1, \dots, \alpha_0 \beta_L), \\
\bpi_g^{t} &\iidsim  \Dir(\alpha \beta_1, \dots, \alpha \beta_g + \kappa, \dots, \alpha \beta_L),  \\
\bmu_g &\iidsim N(\mathbf{0}, \tau^2 I_p), \qquad \sigma_g^2 \iidsim \InvGamma(a/2, b/2), \quad g = 1, \dots, L, \quad t = 2, \dots, T. \\
\end{split}
\end{equation}
Practically, the weak limit approximation transforms an infinite transition matrix into a finite $L \times L$ matrix. The parameter $L$ gives us control over the approximation with the accuracy increasing as $L$ goes to infinity \citep{ishwaran2000}. In practice, one often sets $L$ larger than some a priori upper limit on the number of communities.

\section{Estimation}\label{sec:estimation}

We take a fully Bayesian approach to estimation. In what follows, we describe a Markov chain Monte Carlo (MCMC) method to sample from the posterior of the HDP-LPCM described in Section \ref{sec:models}. We implement a Metropolis-Hastings within Gibbs MCMC scheme with the goal of identifying an evolving community structure consistent with the inferred posterior distribution.

\subsection{Blocked Metropolis-Hastings within Gibbs Sampler}

As outlined in Section \ref{sec:models}, the joint distribution over all the variables (Equation (\ref{eq:prob_factorization})) factors as
\begin{equation}
\underbrace{p(Y_{1:T} \condit \latentpos, \beta_0)}_\text{Equation (\ref{eq:network-likelihood})}
\cdot \overbrace{p(\latents \condit \Pi_{2:T}, \bpi_0^1, \clusterparams, \lambda)}^\text{Equation (\ref{eq:latent-community-model})}
\cdot \underbrace{p(\Pi_{2:T}, \bpi_0^1 \condit \bbeta) \cdot p(\clusterparams) \cdot p(\bbeta)}_\text{Equation (\ref{eq:weak_limit})}.
\end{equation}
To complete the model we assign the following priors:
\begin{equation}
\beta_0 \sim N(\mu_{\beta_0}, \sigma^2_{\beta_0}), \qquad \lambda \sim N_{(0, 1)}(\mu_{\lambda}, \sigma^2_{\lambda}),
\end{equation}
where $N_{(0, 1)}(\mu, \sigma^2)$ indicates the normal distribution with mean $\mu$ and variance $\sigma^2$ truncated to the range $(0, 1)$.

We can realize these samples by following a Gibbs sampling algorithm in which we iteratively sample from the appropriate conditional distributions of $\latentpos$, $\labels$, $\bpi_0^1$, $\Pi_{2:T}$, $\bbeta$, $\clusterparams$, $\lambda$, and $\beta_0$. Due to our choice of priors, most conditional distributions are conjugate so that the Gibbs updates are derived in standard fashion. See Section \ref{subsec:conditionals} of the supplementary materials for the details on these conditional distributions. Metropolis-Hastings (MH) steps are necessary for the latent positions and the intercept. In both cases, we use a normal random walk proposal. In all experiments, we tune the proposal step sizes using an initial tuning phase so that the proposed moves have a  25\% - 40\% acceptance rate. The only samplers that need special care are the block sampler for the actor labels $\mathcal{Z}_{1:T}$, the sampler for the global prior $\bbeta$, as well as the additional hyperparameter sampling schemes. We describe these samplers in the next sections. The full Metropolis-Hastings within Gibbs sampler is outlined in Algorithm \ref{alg:blockgibbs}.

In the following sections, we use dot notation to indicate summations over an index, e.g., for a random variable $w_{ab}$, $w_{\cdot b} = \sum_a w_{ab}$, $w_{a\cdot} = \sum_b w_{ab}$ and $w_{\cdot \cdot} = \sum_a \sum_b w_{ab}$. In addition, we use $n_{kjt}$ to denote the number of actors that transitioned from group $k$ to group $j$ at time $t$ and $n_{kt}$ to denote the number of actors in group $k$ at time $t$.


\begin{sampler}[hp]
Given the previous initial distribution $(\bpi_0^1)^{(\ell - 1)}$, a set of state-specific transition probabilities $\Pi_{2:T}^{(\ell - 1)}$, the global transition distribution $\bbeta^{(\ell-1)}$, group parameters $(\clusterparams)^{(\ell - 1)}$, node labels $\labels^{(\ell-1)}$, latent positions $\latentpos^{(\ell-1)}$, the likelihood specific parameters $\beta_0^{(\ell-1)}$, and blending coefficient $\lambda^{(\ell-1)}$, update the current parameters as follows:

\begin{enumerate}

\item Initialize current parameters to the values of the $(\ell - 1)$th iteration.

\item Update latent positions $\latentpos$ via MH with a normal random walk proposal.

\item Update $\beta_0$ via MH with a normal random walk proposal.

\item Update node labels $\labels$ as in Algorithm \ref{alg:label_sampler}.

\item Sample the auxiliary variables $\boldm, \boldmbar, \boldw$ as in Algorithm \ref{alg:aux_sampler}.

\item Update the global transition distribution by sampling
\begin{equation*}
\bbeta \sim \Dir(\gamma / L + \barm_{\cdot 1 \cdot}, \dots, \gamma / L + \barm_{\cdot L \cdot}).
\end{equation*}

\item For each $k \in \set{1, \dots, L}$, sample a new initial distribution based on the initial assignments:
\begin{equation*}
\bpi_0^1 \sim \Dir(\alpha_0 \beta_1 + n_{011}, \dots, \alpha_0 \beta_L + n_{0L1}).
\end{equation*}

\item For each $k \in \set{1, \dots, L}$ and $t \in \set{2, \dots, T}$, sample a new transition distribution based on the sample assignments:
\begin{equation*}
\bpi_{k}^{t} \sim \Dir(\alpha \beta_1 + n_{k1t}, \dots, \alpha \beta_k + \kappa + n_{kkt}, \dots, \alpha \beta_L + n_{kLt}).
\end{equation*}

\item Update cluster parameters. For each $k \in \set{1, \dots, L}$:
\begin{equation*}
\begin{split}
\bmu_k &\sim N(\boldmubar_k, \barsigma_k^2 I_p), \\
\sigma_k^2 &\sim \InvGamma((n_{k \cdot} p + a)/2, \bar{b}/2),
\end{split}
\end{equation*}

where $\boldmubar_k, \barsigma_k^2$ are defined in Equation (\ref{eq:mu_posterior}) and $\bar{b}$ is defined in Equation (\ref{eq:sigma_posterior}) in Section \ref{subsec:conditionals} of the supplementary materials.

\item Update blending coefficient $\lambda$:
\begin{equation*}
\lambda \sim N_{[0,1]}(\bar{\mu}_{\lambda}, \bar{\sigma}_{\lambda}^2),
\end{equation*}
where $\bar{\mu}_{\lambda}, \bar{\sigma}_{\lambda}^2$ are defined in Equation (\ref{eq:lambda_posterior}) in Section \ref{subsec:conditionals} of the supplementary materials.

\item Update remaining hyperparameters as in Algorithm \ref{alg:hyperparameters} in Section \ref{subsec:hyper_appendix} of the supplementary materials.

\end{enumerate}
\caption{Blocked Metropolis-Hastings within Gibbs sampler for the HDP-LPCM.}
\label{alg:blockgibbs}
\end{sampler}

\subsection{Sampling $\mathcal{Z}_{1:T}$}
Although previous work on dynamic latent space clustering models \citep{sewell2017} update $\mathcal{Z}_{1:T}$ coordinate wise, we chose to utilize a blocked Gibbs sampler. In particular, we use a variant of the forward-backward algorithm for AR-HMMs \citep{rabiner1989} to jointly sample an actor's group assignments $Z_{1:T}^i$. We made this change to increase the mixing rate of the sampler. Previous work has shown that sampling $Z_{1:T}^i$ as a block as opposed to coordinate wise can lead to drastically faster mixing rates in HMMs. This is especially the case when the elements of $Z_{1:T}^i$ are highly related in the posterior distribution, which is likely due to our stickiness assumption \citep{scott2002}. The full sampling algorithm is detailed in Algorithm \ref{alg:label_sampler}. A derivation of the sampler is provided in Section \ref{subsec:label_sampler} of the supplementary materials. A similar sampler is found in \citet{fox2011b}; however, they do not consider multiple state sequences with time-inhomogeneous transition probabilities as we do in this work.

\begin{sampler}[h]
Given transition probabilities $\Pi_{1:T}$, group parameters $\clusterparams$, and latent positions $\latentpos$, for each $i \in \set{1, \dots, n}$, update actor labels as follows:
\begin{enumerate}
    \item Working sequentially backwards in time, calculate messages $m_{t,t-1}(k)$:
    \begin{itemize}
        \item[(a)] For each $k \in \set{1, \dots, L}$, initialize messages to
        \begin{equation*}
        m_{T+1, T}(k) = 1.
        \end{equation*}
        \item[(b)] For each $t \in \set{T, \dots, 2}$ and for each $k \in \set{1, \dots, L}$, compute
        \begin{equation*}
        m_{t, t-1}(k) =
        \sum_{j=1}^L \pi_{kj}^{t} N(\X_t^i \condit \lambda \bmu_j + (1-\lambda)\X_{t-1}^i, \sigma_j^2 I_p) m_{t+1, t}(j), \quad  t > 1.
        \end{equation*}
    \end{itemize}
    \item Sample state assignments $Z_{1:T}^i$ working sequentially forward in time. In addition, update the cluster counts starting with $n_{jkt} = 0$. For $t \in \set{1, \dots, T}$:
    \begin{itemize}
    \item[(a)] For each $k \in \set{1, \dots, L}$, compute the probability
    \begin{equation*}
    f_k(\X_t^i) =
    \begin{cases}
    \pi_{0k}^{1}N(\X_1^i \condit \bmu_{k}, \sigma_k^2 I_p) m_{2, 1}(k), & t = 1, \\
    \pi_{Z_{t-1}^i k}^{t}N(\X_t^i \condit \lambda \bmu_{k} + (1-\lambda)\X_{t-1}^i, \sigma_k^2 I_p)m_{t+1,t}(k), & t > 1.
    \end{cases}
    \end{equation*}
    \item[(b)] Sample a state assignment $Z_t^i$:
    \begin{equation*}
    Z_t^i \sim \sum_{k=1}^L f_k(\X_t^i) \delta(Z_t^i, k).
    \end{equation*}
    \end{itemize}
\end{enumerate}
\caption{Blocked Gibbs sampler for actor labels $Z_{1:T}^i$.}
\label{alg:label_sampler}
\end{sampler}

\subsection{Sampling $\beta$}

We follow the sampling strategy outlined in \citet{teh2006} and \citet{fox2011a}. These samplers use the Chinese restaurant franchise (CRF) metaphor for the hierarchical Dirichlet process to derive closed-form conditional distributions. In the CRF metaphor, each group $j \in \set{1, \dots, J}$ in Equation (\ref{eq:hdp_model}) is associated with a restaurant. The entire collection of restaurants (or franchise) shares a common menu of dishes $\btheta_{1:K}$, where $K$ is the total number of draws from the base measure $H$. The metaphor describes the probabilistic mechanism used by the HDP to partition a collection of individuals (or customers) pre-assigned to $J$ restaurants into $K$ clusters that are identified by their common dish assignments.

For completeness and to properly describe the auxiliary variable samplers in the next section, we present a brief description of the CRF metaphor. As customers arrive at a restaurant, the host seats them at an occupied table with probability proportional to the number of people already seated there and assigns them to a new table with a probability proportional to $\alpha$. When the first customer arrives at a table, a single dish from the global menu is served. The probability to receive an existing dish is proportional to the number of times that dish is served across all restaurants, and a new dish is prepared with probability proportional to $\gamma$. For the sticky HDP, the dish assignment process is modified to account for the stickiness parameter $\kappa$. In this case, each restaurant has a specialty dish. Once the waiter receives a table's order, the waiter has a probability $\rho = \kappa / (\alpha + \kappa)$ to override that order and serve the restaurant's specialty dish instead.

In the HDP-LPCM, we identify each row of the transition matrices, $\set{\bpi_{g}^t}$, with a restaurant in the franchise. In other words, each pair $(g, t)$ indexes the restaurants in the CRF metaphor in a way that a bijective map exists between restaurant $j \mapsto (g, t)$. Furthermore, the customers assigned to restaurant $j$ correspond to all actors assigned to group $g$ at time $t - 1$. With these mappings in mind, we can directly apply the re-sampling algorithm for $\bbeta$ described in \citet{fox2011a} with a few straightforward modifications to account for the multiple hidden state sequences.

\subsection{Sampling $\mathbf{m}, \mathbf{\bar{m}}$, and $\mathbf{w}$}
The sampler for $\bbeta$ and the hyperparameter samplers presented in the next section require the introduction of three auxiliary variables $\mathbf{m}, \mathbf{\bar{m}}$, and $\mathbf{w}$. These variables keep track of certain statistics of the CRF sampling process that when made available preserve conjugacy in the model. The components of these variables are as follows: $m_{jks}$ is the number of tables at restaurant $j$ that ordered dish $k$ at time $s$, $\bar{m}_{jks}$ is the number of tables at restaurant $j$ that were served dish $k$ at time $s$, and $w_{jts}$ is a binary variable indicating whether table $t$ at restaurant $j$ at time $s$ had their dish choice overridden by the waiter. Note that $m_{jks}$ and $\bar{m}_{jks}$ may differ due to the waiter overriding a table's order. The algorithm derived to sampler these auxiliary variables is similar to the one found in \citet{fox2011a} with modifications that account for the increased number of restaurants. We present our auxiliary variable sampler in Algorithm \ref{alg:aux_sampler}. 

\begin{sampler}[h]
Given $n_{jks}$, the number transitions from state $j$ to state $k$ at time $s$, and $\rho = \kappa / (\alpha + \kappa)$, sample the auxiliary variables $\boldm, \boldmbar, \boldw$ as follows:
\begin{enumerate}
\item For each $k \in \set{1, \dots, L}$, set $m_{0k1} = \bar{m}_{0k1} = 0$, and $n = 0$. For each customer in restaurant $0$ eating dish $k$ at time $1$, that is for $i = 1, \dots, n_{0k1}$, sample
\begin{equation*}
x \sim \Bern\left(\frac{\alpha_0 \beta_k}{n + \alpha_0 \beta_k} \right).
\end{equation*}

Increment $n$ by one, and if $x = 1$ increment $m_{0k1}$ and $\bar{m}_{0k1}$ by one.

\item For each $(j, k) \in \set{1, \dots, L}^2$ and $s \in \set{2, \dots, T}$, set $m_{jks} = 0$, and $n = 0$. For each customer in restaurant $j$ eating dish $k$ at time $s$, that is for $i = 1, \dots, n_{jks}$, sample
\begin{equation*}
x \sim \Bern\left(\frac{\alpha \beta_k + \kappa \delta(j, k)}{n + \alpha \beta_k + \kappa \delta(j,k)} \right).
\end{equation*}
Increment $n$ by one, and if $x = 1$ increment $m_{jks}$ by one.

\item For each $j \in \set{1, \dots, L}$ and $s \in \set{2, \dots, T}$, sample the number of overridden dish orders in restaurant $j$ at time $s$:
\begin{equation*}
w_{j\cdot s} \sim \Bin\left(m_{jjs}, \frac{\rho}{\rho + \beta_j(1 - \rho)}\right).
\end{equation*}
Set the number of tables in restaurant $j$ ordering dish $k$ at time $s$ to:
\begin{equation*}
\barm_{jks} =
\begin{cases}
m_{jks}, & j \neq k, \\
m_{jjs} - w_{j\cdot s}, & j = k.
\end{cases}
\end{equation*}
\end{enumerate}

\caption{Sampler for auxiliary variables needed for the update of the global transition distribution $\bbeta$. Note that we do not need to incorporate overridden dishes for $m_{0k1}$ because we draw $\bpi_0^1$ from a $\DP(\alpha_0, \bbeta)$, which has no stickiness parameter $\kappa$.}
\label{alg:aux_sampler}
\end{sampler}


\subsection{Sampling Hyperparameters}
The HDP-LPCM has many hyperparameters that must be set by the practitioner. If one has no a priori knowledge of appropriate hyperparameter settings, it can be advantageous to incorporate hyperparameter exploration into the sampling procedure. In this section, we detail additional hyperparameter samplers as well as our initial choices of hyperparameter values used in the applications of the following sections. The full hyperparameter sampler is presented in Algorithm \ref{alg:hyperparameters} in Section \ref{subsec:hyper_appendix} of the supplementary materials.

In our experience, the hyperparameters that influence inference the most are $\tau^2$, $b$, $\gamma$, $\alpha_0$, $\alpha$, and $\kappa$. The prior variance of the group means $\tau^2$ controls the scale of the latent space. The prior scale of the group variances $b$ controls the size of the clusters in the network. The concentration parameters $\gamma$, $\alpha_0$, and $\alpha$ influence the a priori number of clusters in the network with higher values indicating more groups. Finally, larger values of $\kappa$ result in a priori higher probabilities of self-transitions. With this intuition about the role of these parameters, we proceed to describe our choice of hyperparameter priors and initial values.

Determining the scale of the latent space is a difficult task a priori due to the unobserved nature of the space. For this reason, we place the following priors on $\tau^2$ and $b$:
\begin{equation}
\tau^2 \sim \InvGamma(a_{\tau}/2, b_{\tau}/2), \qquad b \sim \Gamma(c/2, 2/d),
\end{equation}
where $\InvGamma(a, b)$ indicates an Inverse Gamma distribution with shape parameter $a$ and scale parameter $b$, and $\Gamma(c, d)$ represents a Gamma distribution with shape and scale parameters $c$ and $d$ respectively. Under these priors, the updates for the variance of the cluster means $\tau^2$ and the scale parameter for the group variances $b$ are conjugate. We use a heuristic similar to the \texttt{latentnet} package \citep{krivitsky2008} to set the remaining hyperparameter values. In particular, we expect the size of the latent space to grow with the number of actors $n$. For this reason, we chose $a_{\tau}$ and $b_{\tau}$ such that $\tau^2$ has mean $\E{\tau^2} = n^{2/p} / 50$  and standard deviation $4 \E{\tau^2}$, where $n$ is the number of nodes and $p$ is the dimension of the latent space. In addition, we chose $b, c$, and $d$ such that the mode of $\sigma^2$ is $\tau^2$ with a standard deviation of $4 \E{b}$. Furthermore, the initial values for $\tau^2$ and $b$ are set to their prior mean values and $a = 2$. The idea behind this choice  is to set a prior that can generate group shapes ranging from the scale of the entire latent space to a much smaller scale.

To update the hyperparameters of the HDP prior $\gamma, \alpha_0, \alpha$, and $\kappa$, we utilize a straightforward modification of the samplers in \citet{teh2006} and \citet{fox2011a}. Both samplers employ or extend the sampler developed in \citet{escobar1995}, which we describe in Algorithm \ref{alg:escobar_west_sampler} in Section \ref{subsec:hyper_appendix} of the supplementary materials. For all simulations and data applications, we place the following disperse priors on the concentration parameters:
\begin{equation}
\gamma \sim \Gamma(1, 0.1), \quad \alpha_0 \sim \Gamma(1, 1), \quad \alpha + \kappa \sim \Gamma(5, 0.1), \quad \rho \sim \Beta(8, 2).
\end{equation}
Note that under these priors, $\E{\rho} = 0.8$, which matches our large prior belief in self-transitions. The initial values are $\gamma = 1$, $\alpha = \alpha_0 = 1$, and $\kappa = 4$. In our experience, the sampler is insensitive to these initial values.

\subsection{Initialization Scheme}

Due to the high dimensionality of our model, we can greatly reduce the number of iterations required to reach convergence by choosing good initial values for the model parameters. To initialize the latent positions $\mathcal{X}_{1:T}^{(1)}$ and the intercept $\beta_0^{(1)}$, we used the MAP estimate from a short MCMC chain (1,000 iterations) of the model with no clustering of \citet{sewell2015}. Note that we modified their sampler to use the undirected likelihood with the linear predictor given in Equation (\ref{eq:undirected_predictor}). Due to the block sampling of the node labels $\mathcal{Z}_{1:T}$ and the fact that they are sampled before the other clustering parameters, the sampler is not affected by the initial values of the node labels. We initialized the group locations and shapes $(\mu_{1:L}, \sigma^2_{1:L})^{(1)}$ by running a k-means algorithm for longitudinal data \citep{genolini2010} on the initial latent positions with $k = L$ and using the empirical mean and variance estimates of the groups as initial values. We set the initial blending coefficient $\lambda^{(1)} = 0.9$ to encourage early clustering. We sampled the remaining parameters $\bbeta^{(1)}$, $(\bpi_0^1)^{(1)}$, and $\Pi_{2:T}^{(1)}$ from their respective priors. We set the prior parameters on the intercept to $\mu_{\beta_0} = \beta_0^{(1)}$ and $\sigma^2_{\beta_0} = 2$. Finally, we set the prior parameters on the blending coefficient to $\mu_{\lambda} = 0.9$ and $\sigma_{\lambda}^2 = 0.01$.

\subsection{Posterior Summarization}\label{subsec:model_selection}

The selection of a single point estimate that adequately summarizes the full posterior is a challenging and open problem in Bayesian nonparametric statistics. A simple solution uses the posterior mode. However, in Bayesian nonparametric models, the MCMC chain often only visits a single partition once, which makes frequency estimates unreliable. Another approach is to choose the partition corresponding to the MAP (maximum a posteriori) estimate. A downside of the MAP approach is that it does not marginalize over the uncertainty in the partitions. As a result, MAP clustering estimates tend to over-fit the data. An ideal summarization methodology should take into account the clustering uncertainty suggested by the posterior samples.

In this work, we take a decision theoretic approach to posterior summarization. Specifically, we select the partition that minimizes the posterior expectation of an appropriately chosen loss function over dynamic clusterings. In the case of static clustering, two popular choices of loss functions are Binder's loss \citep{binder1978} and the variation of information \citep{meila2007}. Their use for posterior summarization was advocated by \citet{lau2012} and \citet{wade2018} respectively. Note that minimizing Binder's loss is equivalent to maximizing the Rand index, another popular measure of clustering performance. Both loss functions have the desirable property that they are metrics over the space of clusterings. Furthermore, the optimization of the expected losses only depends on the posterior co-occurrence probabilities, so avoids complications due to label-switching.

To extend this approach to dynamic clusterings, we minimize the posterior expected time-averaged variation of information (VI):
\begin{equation}
\argmin_{\hat{\mathcal{Z}}_{1:T}} \, \mathbb{E}\left[\frac{1}{T}\sum_{t=1}^T\text{VI}(\mathcal{Z}_t, \hat{\mathcal{Z}}_t) \; \middle\vert \;  Y_{1:T} \right]
\end{equation}
which averages the static VI's at each time-step. Similar to \citet{wade2018}, this is accomplished by minimizing the following lower-bound:
\begin{equation}
\argmin_{\hat{\mathcal{Z}}_{1:T}}\left\{\sum_{t=1}^T \sum_{i=1}^n \log(\sum_{j=1}^n \ind{\hat{Z}_{t}^j = \hat{Z}_{t}^i}) - 2 \sum_{t=1}^T \sum_{i=1}^n \log(\sum_{j=1}^n p(Z_{t}^j = Z_{t}^i \condit Y_{1:T}) \ind{\hat{Z}_{t}^j = \hat{Z}_{t}^i})\right\}.
\end{equation}
Notice that this expression only depends on the posterior co-occurrence probabilities, which can be estimated from the posterior samples. We perform this optimization by searching over all posterior samples from the MCMC chain (after an appropriate burn-in). Possible ties are broken by selecting the sample with the highest likelihood (Equation (\ref{eq:network-likelihood})). Note that this minimization does not require the partitions to be visited by the Markov chain; however, by restricting the search to the sampled partitions, we have access to the associated latent space for later visualization of the network.

\subsection{Non-Identifiability of the Latent Positions}

Since latent position models depend on the distance between actors, it is clear that they are invariant to translations, rotations, and reflections of the latent space. Any posterior inference that utilizes these positions must correct for such a non-identifiability. We take the approach of Procrustes matching commonly employed in the literature. This involves post-processing the samples by rotating and translating them to match a reference layout. For a detailed description of this procedure, see the original work by \citet{hoff2002}. We use the sample chosen by the procedure in Section \ref{subsec:model_selection} as our reference layout.


\section{Simulation Study} \label{sec:simulation}

We designed a simulation study to assess the HDP-LPCM's performance on synthetic networks with known community dynamics that mimic those found in real-world networks. We considered two scenarios: a sequence of networks with (1) a single static number of groups active over all time steps, and (2) a number of groups that change over time because the groups have either merged or split. We labeled these scenarios the time-homogeneous and time-inhomogeneous simulations, respectively. We expect both scenarios to occur in real-world dynamic networks, so it is important that the HDP-LPCM adequately handles both regimes.

We evaluated the performance of our method in four different ways.  To measure the goodness-of-fit of the HDP-LPCM on the training data, we calculated the in-sample area under the receiver operating characteristic curve (AUC) based on the model's edge predictions. To evaluate the performance of the node level clustering, we used two popular metrics: the adjusted Rand index (ARI) and the variation of information (VI) described in Section \ref{subsec:model_selection}. For the ARI, a value closer to one indicates a good clustering. For the VI, a value closer to zero indicates a good clustering. To measure the HDP-LPCM's ability to quantify uncertainty in the number of clusters, we recorded the posterior probabilities of $\abs{G_t}$ at each time point. To assess the ability of our posterior summarization methodology (Section \ref{subsec:model_selection}) to select the correct number of communities, we recorded the number of active communities selected by the model at each time step and compared to the ground truth value.


All models were estimated using a Markov chain consisting of 50,000 samples. We used the initial 5,000 samples to tune the step sizes of the Metropolis-Hastings samplers. We discarded the following 10,000 samples as burn-in leaving 35,000 samples for inference. Furthermore, we fixed the truncation level of the weak-limit approximation (Equation (\ref{eq:weak_limit})) to $L = 10$ in all experiments. Finally, we repeated each simulation 50 times. For each repetition, we simulated a network starting with a different random seed.

\subsection{Time-Homogeneous Simulation}\label{subsec:time-homogeneous}

This simulation contains a single set of groups and transition matrices common to all time points. We used a simulation scheme similar to the one found in \citet{sewell2017} but modified for undirected networks. The networks were generated according to the sampling mechanisms described in Sections \ref{subsec:distance_models} and \ref{subsec:hmm-network}. We fixed the total number of groups to $G = 6$ at each time point. We chose the blending coefficient $\lambda = 0.8$, the dimension of the latent space $p = 2$, and the intercept $\beta_0 = 1.0$. 
We set the group locations to
\begin{equation}\label{eq:sim1_locs}
\parentheses{\bmu_1, \dots, \bmu_6} =
\begin{pmatrix}
-1.5 & 1.5 & -3 & 3 &  0 & 0 \\
 0   & 0   &  0 & 0 & -2 & 2
\end{pmatrix}.
\end{equation}
We drew the group shapes $\sigma_g$ from a $\InvGamma(6, 0.5 \times 10^{-1})$ distribution, and the initial distribution $\bpi_0^1$ from a six-dimensional $\Dir(10, \dots, 10)$. The six rows of the transition matrix, $\bpi_g^t$ for $g = 1, \dots, 6$ and $t = 2, \dots, 6$, were chosen to be proportional to
\begin{equation}\label{eq:trans_probs}
\begin{split}
\Bigg(&\frac{1}{\norm{\bmu_1 - \bmu_g}}, \dots, \frac{1}{\norm{\bmu_{g-1} - \bmu_g}}, \\
      &\text{const} \times \max_{k \neq g}\left\{\frac{1}{\norm{\bmu_k - \bmu_g}}\right\}, \frac{1}{\norm{\bmu_{g+1} - \bmu_g}},\dots, \frac{1}{\norm{\bmu_G - \bmu_g}} \Bigg).
\end{split}
\end{equation}
We set the constant in the above equation equal to 20, which yields self-transition probabilities ranging from 0.83 to 0.89. The latent actor positions, $\mathcal{X}_{1:T}$, and group assignments, $\mathcal{Z}_{1:T}$, were drawn from Equation (\ref{eq:latent-community-model}). Finally, the adjacency matrices were generated according to Equation (\ref{eq:adjacency_model}) with the linear predictor $\eta_{ijt}$ given by Equation (\ref{eq:undirected_predictor}). We simulated 50 dynamic networks with $T = 6$ time points and $n = 120$ actors each.



\begin{figure}[htbp]
\centering
\includegraphics[width=\textwidth]{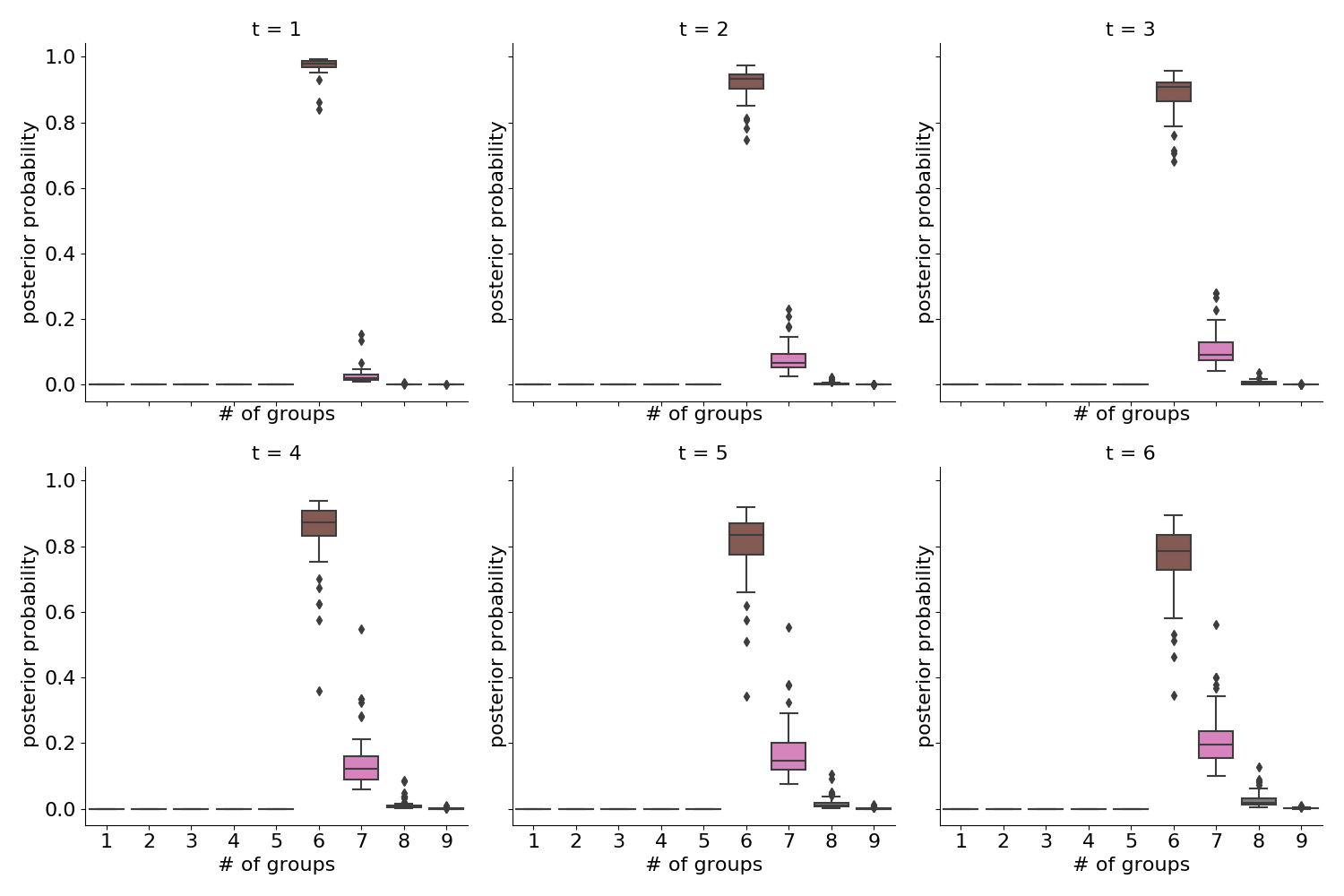}
\caption{Boxplots showing  posterior probabilities of $\abs{G_t}$ over the 50 time-homogeneous simulations. The simulations contained $\abs{G_t} = 6$ communities at each time point. In most cases, the HDP-LPCM assigns high posterior probability to the six community partition.}
\label{fig:homo_post}
\end{figure}

\begin{figure}[htbp]
\centering
\includegraphics[width=0.7\textwidth]{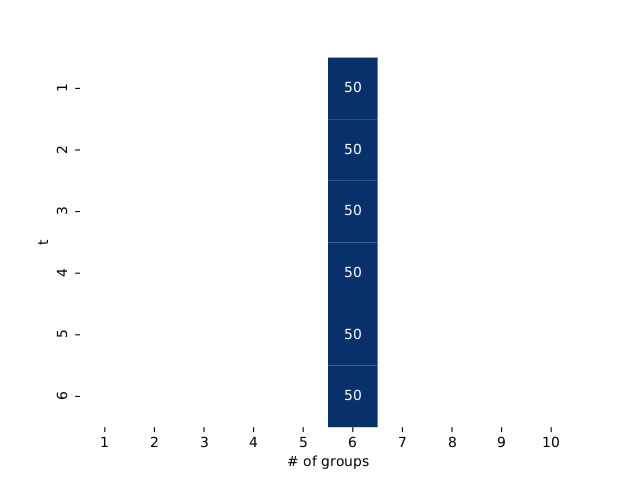}
\caption{The number of groups estimated at each time step by the VI estimator over the 50 time-homogeneous simulations. The data was generated with the  number of clusters $\abs{G_t} = 6$ at all time steps. Our posterior summarization methodology is able to correctly select the six group partition in all simulations.}
\label{fig:homo_counts}
\end{figure}

The results of these simulations are displayed in the first row of Table \ref{tab:performance_metrics}. The in-sample AUC is 0.842, which indicates that the model fits the observed data well. Furthermore, the VI is 0.0324 and the adjusted Rand index is 0.990. Since the values are close to zero and one respectively, we can conclude that the clustering performance is very good. Next, we evaluate the model's ability to quantify the uncertainty in the number of clusters. Boxplots of the posterior probabilities of $\abs{G_t}$ over the 50 simulations are displayed in Figure \ref{fig:homo_post}. In most cases, $\abs{G_t} = 6$ is correctly identified with the highest posterior probability. Finally, we assess the ability of our posterior summarization method to select a partition with the correct number of communities. The number of groups selected by this method at each time step is displayed in Figure \ref{fig:homo_counts}. Our methodology is able to correctly select the six group partitioning in all fifty simulations.


\begin{table}[htbp]
\centering
\begin{tabular}{@{}llll@{}} \toprule
Simulation & AUC & Average VI & Average ARI \\
\bottomrule[\lightrulewidth]
Time Homogeneous & 0.842 (0.010) & 0.0324 (0.0240) & 0.990 (0.008) \\
Time Inhomogeneous & 0.850  (0.005) & 0.0292 (0.338) & 0.990 (0.184) \\
\bottomrule
\end{tabular}
\caption{Median performance metrics over all 50 time-homogeneous / time-inhomogeneous simulations. Standard deviations are displayed in parentheses.}
\label{tab:performance_metrics}
\end{table}

\subsection{Time-Inhomogeneous Simulation}\label{subsec:inhomogeneous_sim}

We designed this simulation to test our model's ability to detect changes in group structure by allowing the number of groups to vary over time. We generated 50 networks with $T = 9$ time points and $n = 120$ actors each. There are $G = 6$ groups in the networks overall. The simulations begin with two groups, these two groups split into six groups at $t = 4$, and then the six groups merge into four groups at $t = 7$. Note that the adjacency matrices generated from this procedure are displayed in the introductory example (Figure \ref{fig:evolving_communities}).


Once again, the networks were simulated according to the sampling mechanisms described in Sections \ref{subsec:distance_models} and \ref{subsec:hmm-network}. This simulation differs from the time-homogeneous simulation in how we specified the transition matrices $\bpi_g^t$. We drew the initial distribution $\bpi_0^1$ from a two-dimensional $\Dir(10, 10)$. We chose the rows of the transition matrix, $\bpi_g^t$ for $g \in G_{t-1}$ and $t = 2, \dots, 9$, proportional to
\begin{equation}\label{eq:sim2_trans_probas}
\begin{split}
\Bigg(&\frac{\ind{1 \in G_t}}{\norm{\bmu_1 - \bmu_g}}, \dots, \frac{\ind{g-1 \in G_t}}{\norm{\bmu_{g-1} - \bmu_g}}, \\
      &\text{const} \times \max_{k \neq g \, : \, k\in G_t}\left\{\frac{1}{\norm{\bmu_k - \bmu_g}}\right\} \ind{g \in G_t},
      \frac{\ind{g+1 \in G_t}}{\norm{\bmu_{g+1} - \bmu_g}},\dots, \frac{\ind{G \in G_t}}{\norm{\bmu_G - \bmu_g}} \Bigg),
\end{split}
\end{equation}
where $G_t$ indicates the set of active groups at time $t$. At $t = 1,2,3$ there are two groups so that $G_{1} = G_{2} = G_{3} = \set{1, 2}$. From $t = 4,5,6$ these two groups split into six groups so that $G_4 = G_5 = G_6 = \set{1, 2, 3, 4, 5, 6}$. Finally, these six groups merge into four groups at $t = 7,8,9$ so that  $G_7 = G_8 = G_9 = \set{3, 4, 5, 6}$. For all time points, we set the constant in Equation (\ref{eq:sim2_trans_probas}) to 20 except for $t = 4$. At $t = 4$, we set the constant equal to 1 so that the nodes were more evenly distributed among the six newly created groups.

\begin{figure}[htbp]
    \centering
    \includegraphics[width=0.6\textwidth]{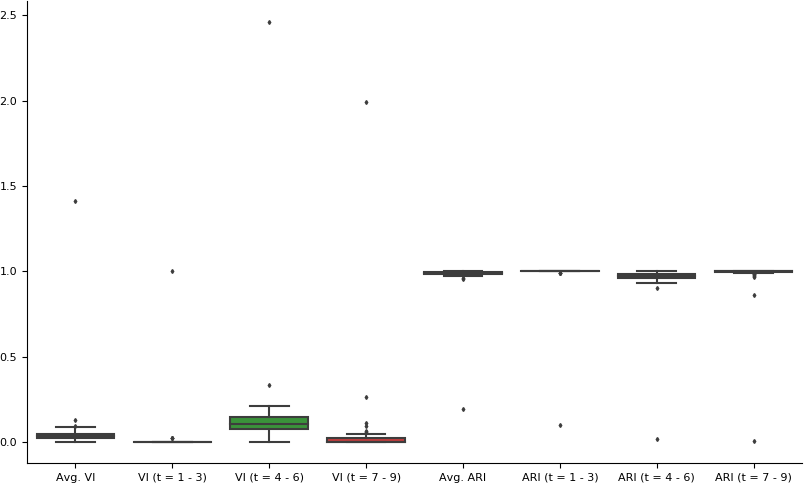}
    \caption{Boxplot of metrics for the time-inhomogeneous simulation. Avg. VI and Avg. ARI indicate average values taken over all times. Other values correspond to the metrics calculated in the interval in the parentheses. For example, VI ($t = 1 - 3$) indicates the variation of information calculated with labels estimated during $t = 1, 2$, and 3.}
    \label{fig:inhomogeneous-metrics}
\end{figure}

The results of the simulation are displayed in the second row of Table \ref{tab:performance_metrics}. The median in-sample AUC is 0.850, which indicates that the model fits the observed data well. Furthermore, the median VI is 0.0292 and the median ARI is 0.990. Since the values are close to zero and one, respectively, we conclude that the clustering performance is very good. We note that the standard deviations of these metrics are much higher in this simulation. To make sense of this discrepancy, we display boxplots of the metrics in Figure \ref{fig:inhomogeneous-metrics}. We see that the VI and the ARI have a few large outliers for some simulations; however, the majority of simulations are clustered near the mean values, which indicates a good clustering.

\begin{figure}[htbp]
\centering
\includegraphics[width=0.75\textwidth]{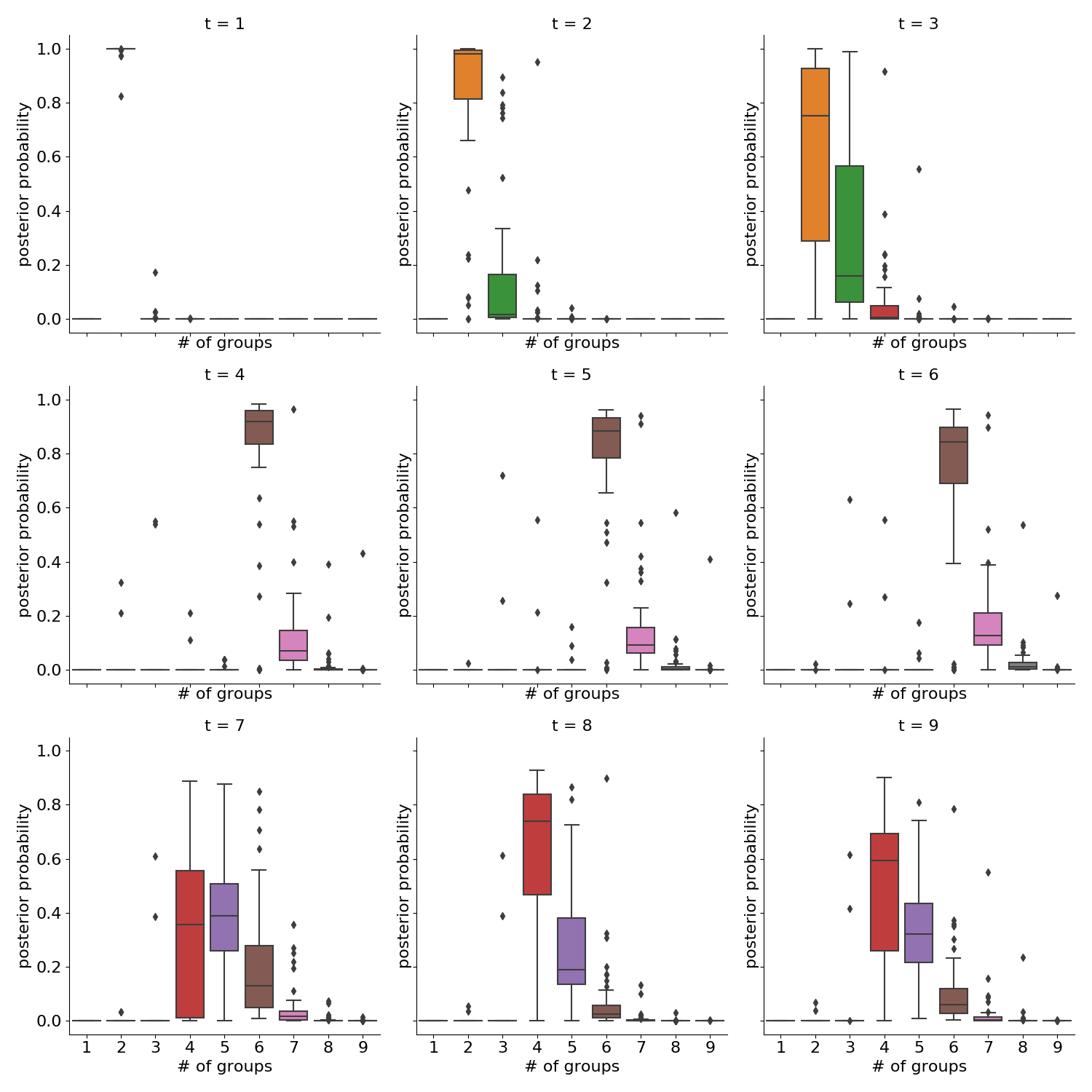}
\caption{Boxplots of the  posterior probabilities of $\abs{G_t}$ over the 50 time-inhomogeneous simulations. The data was generated with $\abs{G_t} = 2$ for $t = 1 - 3$, $\abs{G_t} = 6$ for $t = 4-6$, and $\abs{G_t} = 4$ for $t = 7-9$. In most cases, the HDP-LPCM is able to assign high posterior probability to the correct number of communities.}
\label{fig:inhomogeneous-posteriors}
\end{figure}

\begin{figure}[htbp]
\centering
\includegraphics[width=0.6\textwidth]{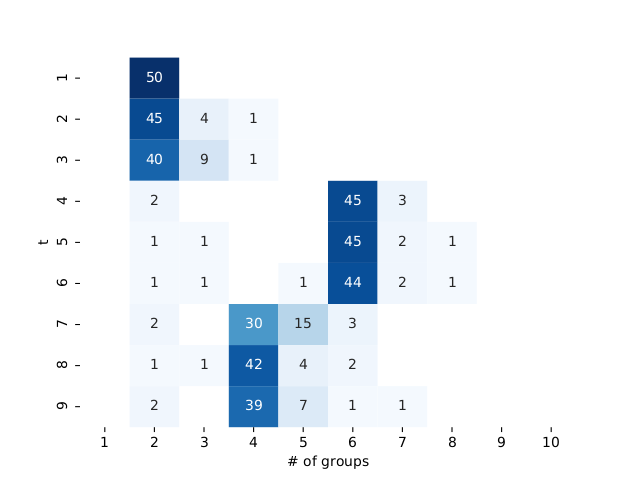}
\caption{The number of groups predicted at each time step by the VI estimator for the time-inhomogeneous simulation. The data was generated with $\abs{G_t} = 2$ for $t = 1 - 3$, $\abs{G_t} = 6$ for $t = 4-6$, and $\abs{G_t} = 4$ for $t = 7-9$. Our posterior summarization method is able to select the correct number of communities in the majority of the simulations.}
\label{fig:inhomogeneous-counts}
\end{figure}

Next, we evaluate the ability of the HDP-LPCM to quantify the changing macro-level community structure. The posterior probabilities for $\abs{G_t}$ are displayed in Figure \ref{fig:inhomogeneous-posteriors}. Recall that the ground truth is $\abs{G_t} = 2$ for $t = 1 - 3$, $\abs{G_t} = 6$ for $t = 4-6$, and $\abs{G_t} = 4$ for $t = 7-9$. For most time points the correct number of groups has the highest posterior probability. The exception is $t = 7$, which is when the six groups merge into four groups. As is common in Dirichlet process mixture models, the model tends to create a few small clusters, which can inflate the group count. This is the case here, where the extra clusters at $t = 7$ tend to be composed of less than five nodes. Once again we assess our posterior summarization method's ability to select a clustering with the correct number of communities. The number of groups selected by this method at each time step is displayed in Figure \ref{fig:inhomogeneous-counts}. For the majority of simulations, our method selects the correct number of clusters. The method struggles the most at $t = 7$, typically because of a few small clusters. Overall, the clustering is correct and we feel confident the HDP-LPCM is able to adequately infer evolving communities.

\section{Real Data Application}\label{sec:real_data}

In this section, we demonstrate the utility of our proposed HDP-LPCM on a variety of real-world dynamic networks with an evolving community structure. We include two applications: inferring changing international military alliances during the Cold War and detecting dynamic plotlines in the television series Game of Thrones. Also, in Section \ref{subsec:sampson} of the supplementary materials, we show that the HDP-LPCM corroborates many previous findings concerning Sampson's monastery network \citep{sampson}, a standard pedagogical example in the network literature.

\subsection{International Military Alliances}

We begin by using the HDP-LPCM to examine changes in international military alliances during the first three decades of the Cold War (1950 - 1979). We use the Formal Alliances (v4.1) dataset curated as part of the Correlates of War Project \citep{gibler2009}. The raw dataset records all formal alliances -- mutual defense pacts, non-aggression treaties, and ententes -- among nations between 1816 and 2012. The goal of our analysis is to uncover the competing political blocs that defined the Cold War period in history and to determine any points in time where that alliance structure changed.

For this analysis, we use the yearly dyadic dataset, which records an undirected edge between two nations if there is a formal alliance between them during that year. To simplify the analysis, we discretized the dynamic networks into five year chunks from 1950 - 1979. We removed the nodes with a degree less than two from each network to focus on the larger political blocs found in the dataset. In addition, we required a nation to have at least one alliance during 1950 - 1979. We binarized the relations so that a connection between nations $i$ and $j$ at time $t$ means they had at least one active alliance during those five years. This preprocessing resulted in $T = 6$ undirected binary networks that each contains $n = 107$ actors.

We fit our proposed model with a truncation level $L = 25$ to the international military alliances networks using 50,000 iterations for tuning, 50,000 iterations for burn-in, which left a remaining 400,000 iterations for inference. The trace plots, autocorrelation functions (ACFs), and marginal densities are displayed in Figure \ref{fig:alliances_traces} in Section \ref{subsec:additional_figures} of the supplementary materials. Visual inspection of the trace plots indicates the algorithm has converged.

\begin{figure}[htbp]
\centering
\includegraphics[height=0.25\textheight]{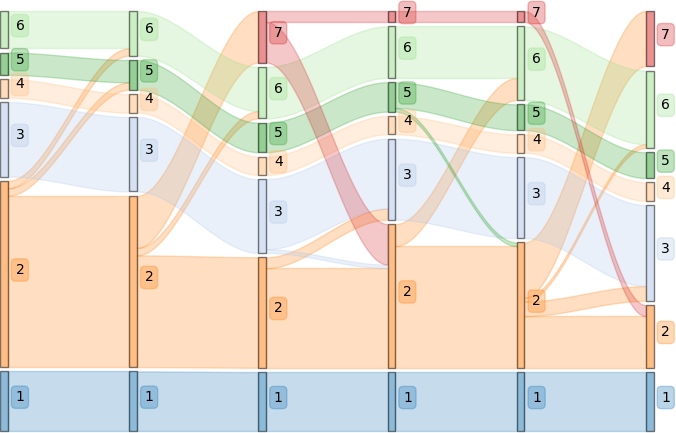}
\caption{Alluvial diagram for the international military alliances network. Each line represents the flux of nodes going from one group to the next at time $t$ to $t+1$. The thickness of the lines is proportional to the number of nodes and the total height represents all nodes.}
\label{fig:alliances_alluvial}
\end{figure}

\begin{figure}[htbp]
    \includegraphics[width=\textwidth]{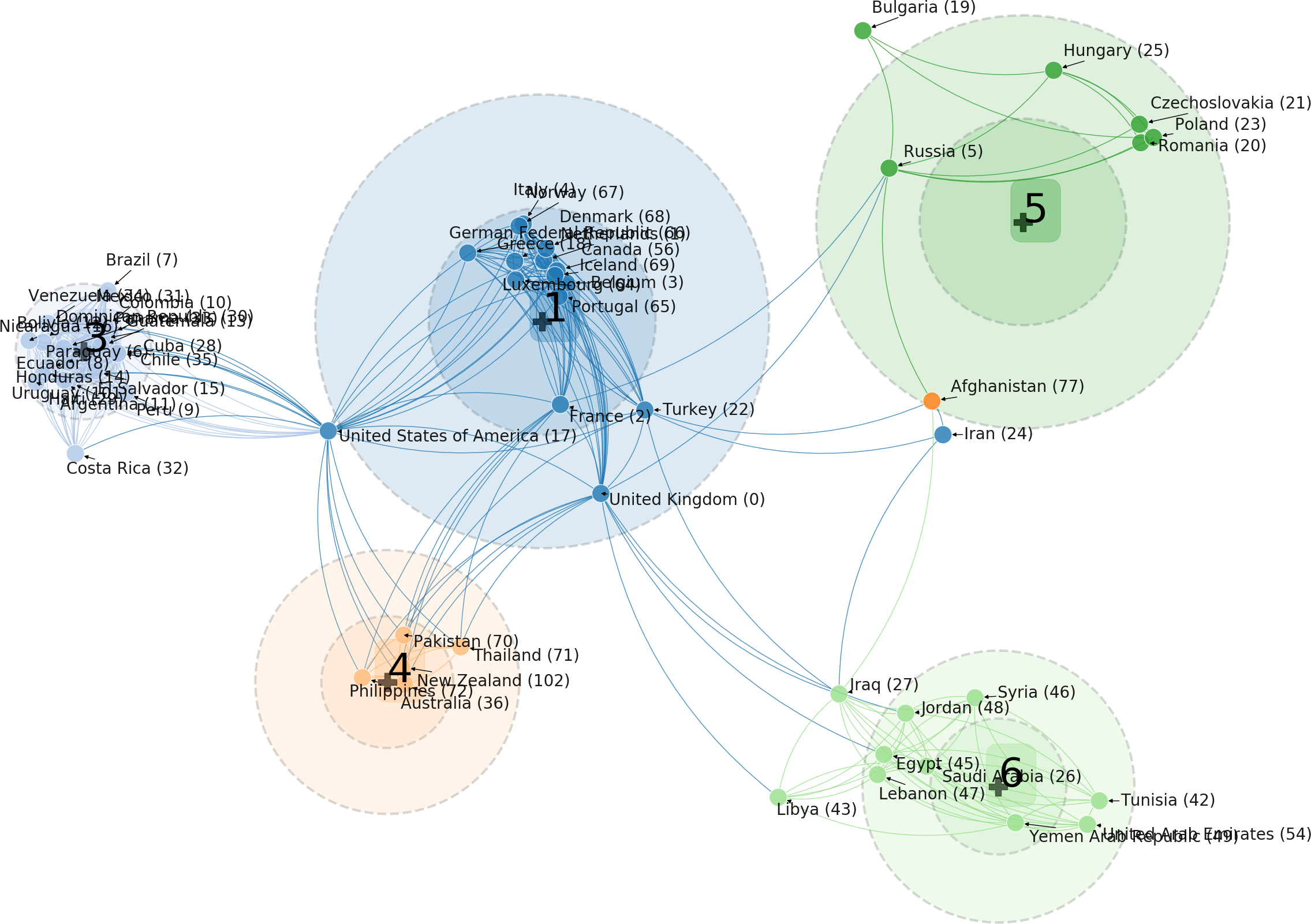}
    \caption{Latent space of international military alliances for the years 1950 - 1954. The group means $\bmu_g$ are denoted by a $+$, and the group shapes $\sigma_g$ are displayed as two-standard deviation ellipses. The names of each nation and node number are annotated. The undirected edges are also displayed. For clarity, all unconnected nodes and group 2's two-standard deviation ellipses are removed.}
    \label{fig:alliances_t0}
\end{figure}

The alluvial diagram (Figure \ref{fig:alliances_alluvial}) of the partition selected with the procedure described in Section \ref{subsec:model_selection} shows that the HDP-LPCM estimates six communities overall with five communities active during all six time points. Note that the model uses group 2 to collect isolated nodes and small intermittent alliances. This interpretation is supported by the fact that $\hat{\sigma}_2 = 132$, which is 28 times larger than the second largest group shape $\hat{\sigma}_1 = 4.67$, and it encompasses most of the latent space. Thus we consider groups 1, 3, 4, 5, 6, and 7 as the major blocs inferred by the model. Note that these blocs strongly dictate the network's dynamics with an inferred blending coefficient $\lambda = 0.994$. The estimated latent spaces for the years 1950 - 1954 and 1960 - 1964 are depicted in Figure \ref{fig:alliances_t0} and Figure \ref{fig:alliances_t2}. The latent spaces for the remaining years are included in the supplementary materials.

\begin{figure}[htbp]
    \includegraphics[width=\textwidth]{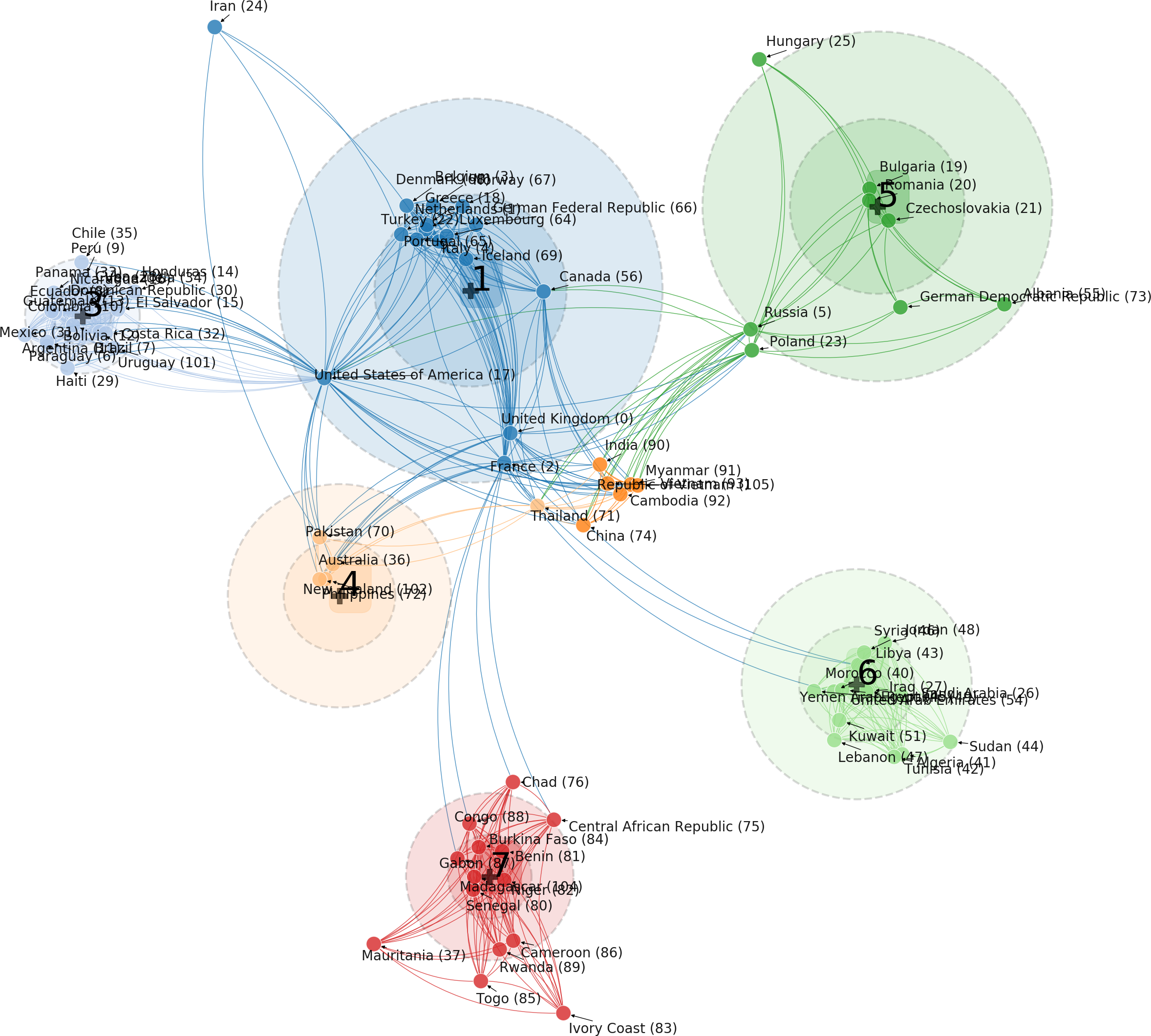}
    \caption{Latent space of international military alliances for the years 1960 - 1964. The group means $\bmu_g$ are denoted by a $+$, and the group shapes $\sigma_g$ are displayed as two-standard deviation ellipses. The names of each nation and node number are annotated. The undirected edges are also displayed. For clarity, all unconnected nodes and group 2's two-standard deviation ellipses are removed.}
    \label{fig:alliances_t2}
\end{figure}

The static communities (groups 1, 3, 4, 5, and 6) coincide with long term regional alliances during the Cold War. Group 1 corresponds to the Western Bloc, consisting primarily of nations that are a part of the North Atlantic Treaty Organization (NATO) and the Western European Union (WEU). The competing Eastern Bloc is represented by group 5, which consisting of the Soviet Union and its satellite states such as East Germany, Czechoslovakia, and Poland. Group 3 consists of the Latin American coalition of the Organization of American States (OAS), founded in 1948 to oppose socialism. Group 4 consists of member nations of the Southeast Asia Treaty Organization (SEATO), which is the Asian equivalent of NATO. Group 6 is the Arab League (at the time the League of Arab States) formed in 1945 to protect the interests of Arab countries. As depicted in the alluvial diagram, an interesting finding of the HDP-LPCM is that there is little exchange of nations between these groups over this time period.
This finding bolsters the claim that ``once the `cold war' confrontation became apparent ... many nations cast their lot with either the American or Soviet bloc" \citep{small1969}.

The evolving community structure (birth of group 7) is a result of the emergence of Africa as a world power. In particular, seventeen African nations gained their independence in 1960 alone. As a result of these newly independent nations, a large number of regional alliances formed in the early 1960s \citep{gibler2009chp}. This is reflected in the HDP-LPCM by the emergence of group 7 at $t = 3$, which encompasses the newly independent African nations. Fewer African alliances formed over the next decade. Our model reflects this fact by only including the Union of Central African States, which consists of former French colonies, in group 7 at $t = 4$ and 5. Finally, the Economic Community of West African States (ECOWAS) formed in 1975, which resulted in the re-introduction of many western African states into group 7 at $t = 6$. The dynamic nature of the HDP-LPCM is essential in revealing the emergence of Africa as a world power and demonstrates the importance of incorporating community evolution in latent space network modeling.


\subsection{Character Interactions in Game of Thrones}

In this section, we study the networks of character interactions in the television series Game of Thrones. The goal of our analysis is to use community detection to pinpoint coherent dynamic plotlines within the series. This is an interesting case study because the series' narrative is known for its many dynamic stories and characters who freely move between them. As such, we expect various groups of characters to form and die out across the series' lifetime and actors to freely move between these groups.

We utilize the networks compiled by \citet{beveridge2018}, who parsed fan-generated scripts found on the user curation site Genius. The original dataset consists of weighted character-character interactions split up between the eight seasons of the television series. The weight of an edge equals the number of interactions during a given season. Since this is primarily a pedagogical example, we restrict the dataset to the first four seasons of the show. To remove minor characters, we only keep interactions that occur greater than or equal to 10 times each season. The final result is $T = 4$ binary undirected networks with a total of $n = 165$ actors each.



We fit the HDP-LPCM to this dataset with a truncation level of $L = 25$ using 50,000 iterations for tuning, 50,000 iterations for burn-in, which left a remaining 400,000 iterations for inference. The trace plots of the unnormalized log-posterior, the intercept $\beta_0$, and the blending coefficient $\lambda$ are displayed in Figure \ref{fig:got_traces} in Section \ref{subsec:additional_figures} of the supplementary materials. Visual inspection of the trace plots indicates the model has converged.

\begin{figure}[htbp]
\centering
\includegraphics[height=0.25\textheight]{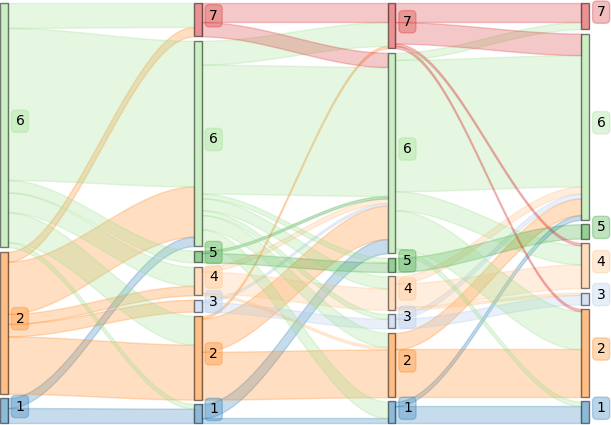}
\caption{Alluvial diagram for the Game of Thrones character interaction networks. Each line represents the flux of nodes going from one group to the next at time $t$ to $t+1$. The thickness of the lines is proportional to the number of nodes and the total height represents all nodes.}
\label{fig:got_alluvial}
\end{figure}

\begin{figure}[hp]
    \centering
    \begin{subfigure}[b]{0.8\textwidth}
        \includegraphics[width=\textwidth]{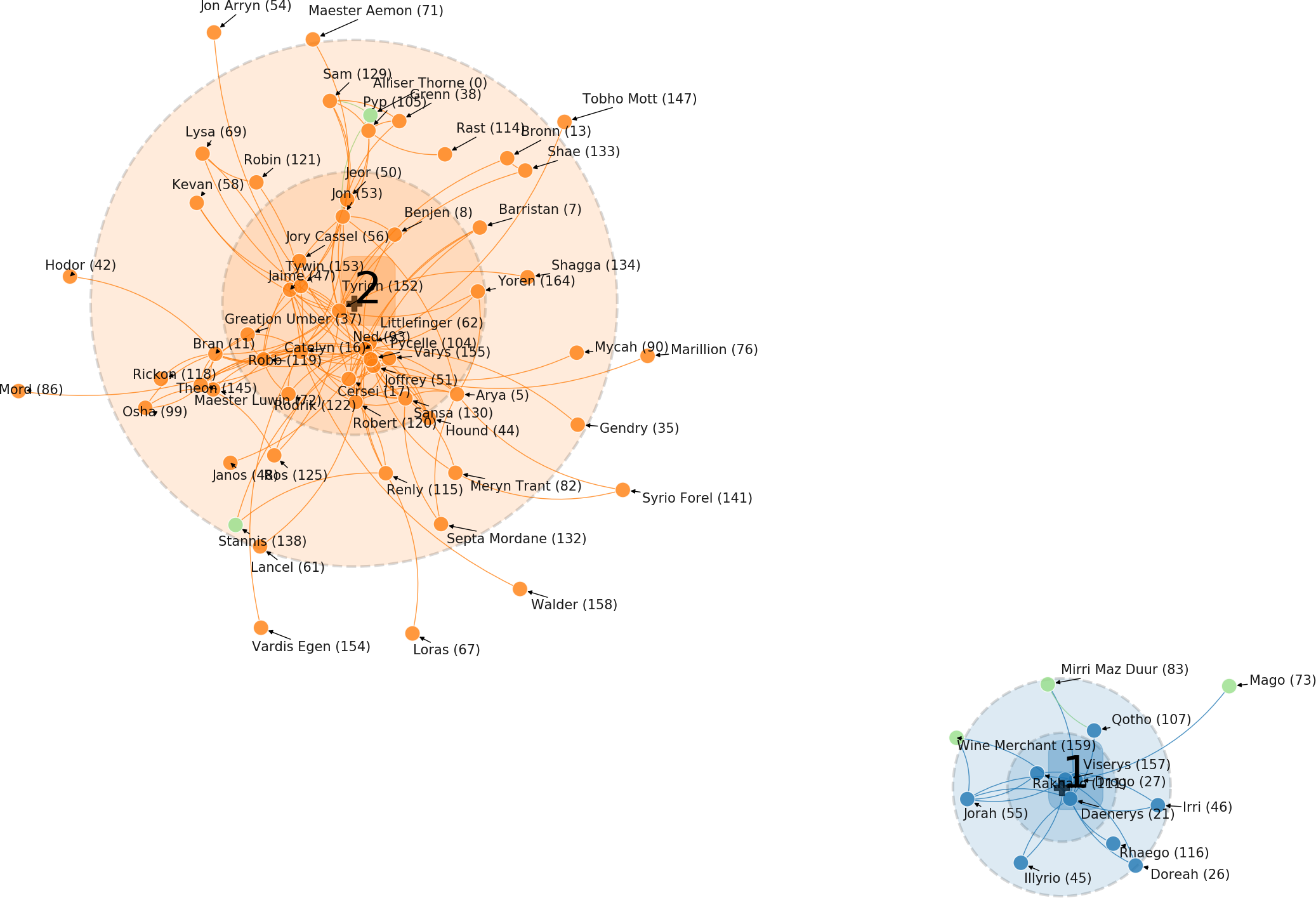}
        \caption*{Season 1}
        \vspace{2em}
    \end{subfigure}
    \hfill
    \begin{subfigure}[b]{0.8\textwidth}
        \includegraphics[width=\textwidth]{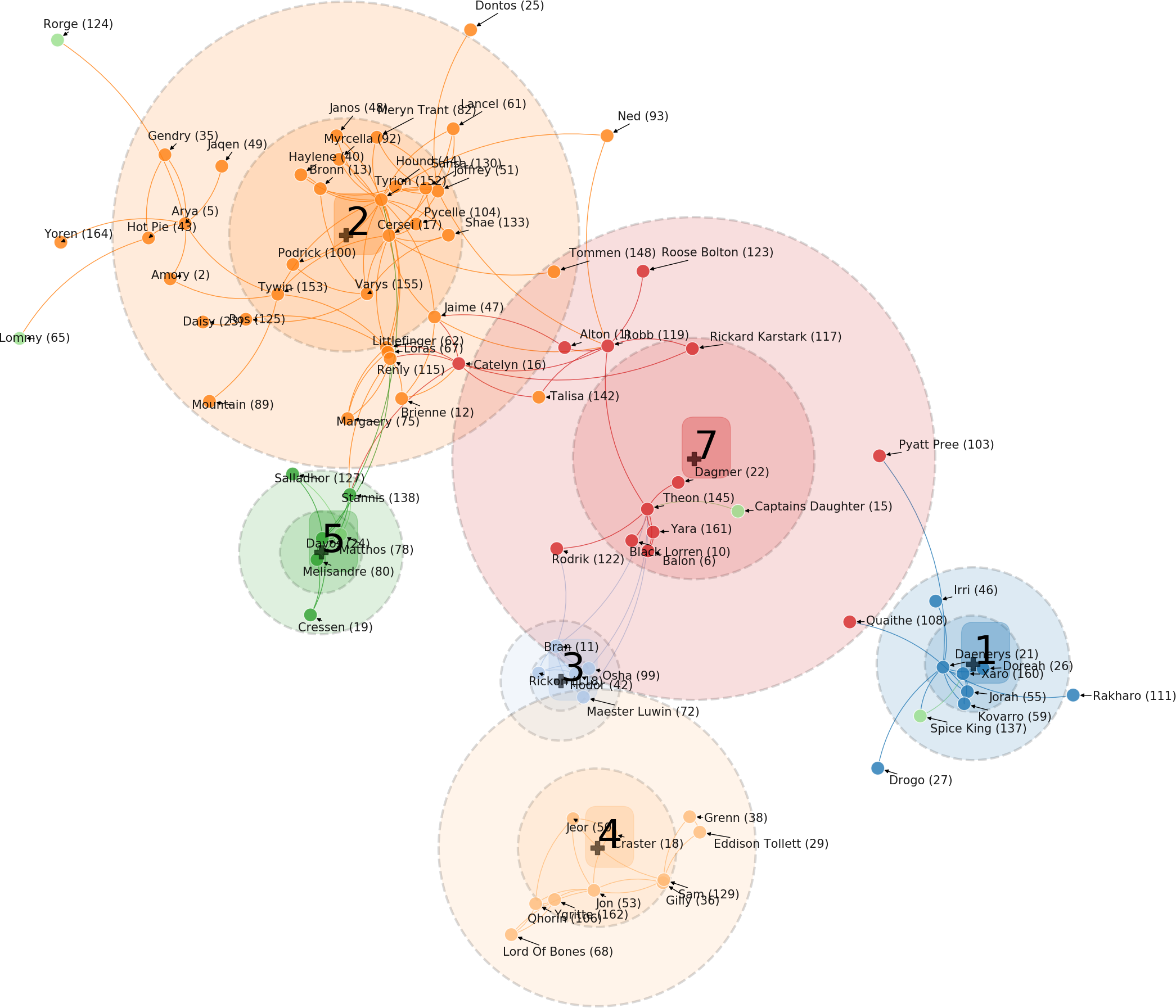}
        \caption*{Season 2}
    \end{subfigure}
    \caption{Latent space of the Game of Thrones character interaction network for season 1 and season 2. The group means $\bmu_g$ are denoted by a $+$, and the group shapes $\sigma_g$ are displayed as two-standard deviation ellipses. The names of each character and node number are annotated. The undirected edges are also displayed. For clarity, all unconnected nodes and group 6's two-standard deviation ellipses are removed.}
    \label{fig:got_latent_space}
\end{figure}

The alluvial diagram (Figure \ref{fig:got_alluvial}) and associated latent space (Figure \ref{fig:got_latent_space}) reveal a dynamic group structure. The model infers six groups overall; however, only three groups are active during the first season of the show. During the second season, group 2 splits off into groups 3, 4, 6, and 7. In addition, group 5 is created out of characters from group 6. Note that the model uses group 6 to collect inactive characters, so we exclude it from further analysis. After the change in season two, the network's group structure remains constant for seasons three and four. The latent space of seasons three and four are included in the supplementary materials. Furthermore, since the blending coefficient $\lambda = 0.974$, we conclude that the groups drive the evolution of the network. Overall, the ability of the HDP-LPCM to infer an evolving group structure is crucial for properly understanding the network's dynamics.

The inferred groups and their dynamics are consistent with the storylines in the Game of Thrones series. During season one, there are two active groups. Group 2 consists of all characters on Westeros, while group 1 centers around Deanery Targaryen's story on Essos. Starting at season two, group 2 contains characters revolving around Arya Stark's story arc, House Lannister, and other characters at King's Landing. Group 3 contains Bran Stark's group of companions (Hodor, Rickon, Osha, Meera, and Jojen) as they travel north of the wall. The plotlines related to the remainder of House Stark are contained in group 7. This group is composed of  both Robb Stark's contingent as well as those related to Theon Greyjoy, who currently rules Winterfell the ancestral home of the Starks. Group 4 pertains to the story north of the wall. It most notably contains Jon Snow and Sam Tarly. Lastly, group 5 revolves around the story of Stannis Baratheon and his quest to regain the Iron throne, which was introduced in season two. In conclusion, our model recovers narratively coherent groups within the Game of Thrones character interaction networks and identifies their temporal dynamics within the storyline.

\section{Discussion}\label{sec:disc}

In this article, we proposed the hierarchical Dirichlet process latent position clustering model (HDP-LPCM) for dynamic networks. This is the first, to our knowledge, latent position model that can detect evolving community structures. In addition, we demonstrated that that HDP-LPCM still performs well when there is a static group structure. To accomplish our modeling goals, we used Bayesian nonparametric priors to provide simultaneous inference over the number of communities at each time point in the network and the dynamics of each actor's latent position. Furthermore, our MCMC inference procedure has computational advantages over existing approaches. In particular, we avoided the BIC approximations of \citet{handcock2007} and the computationally expensive process of estimating a large number of models.

In this work, we focused solely on community detection for binary undirected networks; however, our methodology can easily be extended to other network types. For example, both directed and weighted networks can be accommodated through the minor modifications of the likelihood presented in \citet{sewell2016weights}. Due to the conditional dependence structure of the HDP-LPCM, one only needs to modify the Metropolis-Hastings steps (i.e., steps 2 and 3) in Algorithm \ref{alg:blockgibbs} to use the new network likelihood.

Common to most latent distance models, our proposed MCMC estimation method is time-intensive for large networks due to the quadratic scaling of the procedure. To alleviate this issue, one could utilize the case-control likelihood of \citet{raftery2012}. This method approximates the full likelihood by sub-sampling the unconnected edges, which allows for inference that scales linearly with the number of nodes in the network. To further decrease the model's runtime, one could create a variational Bayes or stochastic variational Bayes \citep{hoffman2013} algorithm. However, the lack of conjugacy in the distance model means such an algorithm is non-trivial to implement. For this reason, we leave these extensions to future work.

Although our model can infer the number of communities from the data, it does not provide inference for the dimension of the latent space $p$. In the numerical studies, we fixed $p = 2$, which allowed for intuitive visualizations as well as being adequate for the networks we considered in this work. Furthermore, we believe such visualizations allow for human-in-the-loop model scrutiny, which is a primary strength of the latent position model over other modeling approaches such as latent feature models \citep{miller2009}. Of course, selecting an appropriate dimension of the latent space is crucial if one is focused on tasks such as link prediction. A straightforward approach is to use a model selection criteria such as BIC or the Watanabe-Akaike information criterion (WAIC) \citep{watanabe2010}. Another possibility is to put a prior over the dimensionality such as in \citet{durante2014}. Regardless, the selection of $p$ in latent space models remains an interesting and open question for future research.

Despite these possible extensions, we believe the HDP-LPCM has a broad range of applicability. As demonstrated on the real-world networks in this paper, applications include the  detection of changing alliances structures as well as inference regarding narrative plotlines. We believe our methodology can be built upon to provide new tools for understanding the sequential evolution of communities in dynamic networks.

\bibliographystyle{apalike}
\bibliography{reference}

\clearpage

\clearpage
\baselineskip=18pt

\setcounter{page}{1}
\setcounter{section}{0}
\setcounter{equation}{0}
\setcounter{table}{0}
\setcounter{figure}{0}
\setcounter{sampler}{0}
\pagenumbering{arabic}
\renewcommand{\thesubsection}{S.\arabic{subsection}}
\renewcommand\theequation{S.\arabic{equation}}
\renewcommand\thetable{S.\arabic{table}}
\renewcommand\thefigure{S.\arabic{figure}}
\renewcommand\thesampler{S.\arabic{sampler}}

\begin{center}
{\Large\textbf{Supplementary Materials for \\
``A Bayesian Nonparametric Latent Space Approach to Modeling Evolving Communities in Dynamic Networks"}} \\

\smallskip

\if0\blind
{
{\large Joshua Daniel Loyal and Yuguo Chen}
}\fi
\if1\blind
{
}\fi
\end{center}


\subsection{Conditional Distributions Used in the Gibbs Sampler}\label{subsec:conditionals}

In this section, we present the conditional distributions used in the Metropolis-Hastings within Gibbs sampler for the HDP-LPCM. In what follows, let $I_{kt} = \set{i : Z_{t}^i = k}$ be the set of nodes in group $k$ at time $t$. Furthermore, let $n_{kt} = \abs{I_{kt}}$ be the number of actors in group $k$ at time $t$, and let $n_{k \cdot} = \sum_{t=1}^T n_{k t}$. We denote conditioning on everything but the parameters of interest as $\condit \cdot$. The full conditionals are
\begin{align}
\begin{split}\label{eq:mu_posterior}
\bmu_k \condit \cdot &\sim  N(\bar{\bmu_k}, \bar{\sigma}^2_k I_p) \quad \text{ where }\\
&\quad \bar{\sigma}_k^2 = \left[\frac{1}{\sigma_k^2}\left(n_{k1} + \lambda^2 \sum_{t=2}^T n_{kt} \right) + \frac{1}{\tau^2}\right]^{-1}, \\
&\quad \bar{\bmu}_k = \bar{\sigma}_k^2\left(\frac{1}{\sigma_k^2} \sum_{i \in I_{k1}} \X_1^i  + \frac{\lambda}{\sigma_k^2}\sum_{t=2}^T \sum_{i \in I_{kt}} (\X_t^i - (1 - \lambda) \X_{t-1}^i) + \frac{1}{\tau^2} \bmu_0 \right),
\end{split}\\[2em]
\begin{split}\label{eq:sigma_posterior}
\sigma_k^2 \condit \cdot &\sim \InvGamma\left((n_{k \cdot} p + a)/2, \bar{b}/2\right) \quad \text{where} \\
&\bar{b} = b + \sum_{i \in I_{k1}} \norm{\X_1^i - \bmu_k}_2^2 + \sum_{t=2}^T \sum_{i \in I_{kt}} \norm{\X_t^i - (1 - \lambda) \X_{t-1}^i - \lambda \bmu_k}_2^2,
\end{split}\\[2em]
\begin{split}\label{eq:lambda_posterior}
\lambda \condit \cdot &\sim N_{(0,1)}(\bar{\mu}_{\lambda}, \bar{\sigma}_{\lambda}^2) \quad \text{where }\\
&\bar{\mu}_{\lambda} = \frac{\mu_{\lambda} + \sigma_{\lambda}^2 \sum_{i=1}^n \sum_{t\geq 2} \frac{1}{\sigma_{Z_t^i}^2} (\X_t^i - \X_{t-1}^i)^T(\bmu_{Z_t^i} - \X_{t-1}^i)}{1 + \sigma_{\lambda}^2\sum_{i=1}^n\sum_{t\geq 2} \frac{1}{\sigma_{Z_t^i}^2} \norm{\bmu_{Z_t^i} - \X_{t-1}^i}_2^2}, \\
&\bar{\sigma}_{\lambda}^2 = \left(\frac{1}{\sigma_{\lambda}^2}  + \sum_{i=1}^n\sum_{t\geq 2} \frac{1}{\sigma_{Z_t^i}^2} \norm{\bmu_{Z_t^i} - \X_{t-1}^i}_2^2\right)^{-1},
\end{split}\\[2em]\label{eq:tau_posterior}
\tau^2 \condit \cdot &\sim \InvGamma((a_{\tau} + L) / 2, (b_{\tau} + \sum_{g=1}^L \norm{\bmu_g}_2^2) / 2), \\[2em]\label{eq:b_posterior}
b \condit \cdot &\sim \Gamma((c + L \, a )/2 , 2 (d + \sum_{g=1}^L \sigma_g^{-2})^{-1}).
\end{align}
The log of the full conditional distributions used in the MH updates of the latent positions $\mathcal{X}_{1:T}$ are
\begin{enumerate}
\item[(a)] for $t = 1$:
\begin{equation}
\begin{split}
\log(p(\X_1^i \condit \cdot)) &\propto \log(p(Y_1 \condit \mathcal{X}_1, \beta_0)) - \frac{1}{2 \sigma_{Z_1^i}^2} \norm{\X_1^i - \bmu_{Z_1^i}}_2^2 \\
&-\frac{1}{2 \sigma_{Z_2^i}^2} \norm{\X_2^i - \lambda \bmu_{Z_2^i} - (1 - \lambda) \X_1^i}_2^2,
\end{split}
\end{equation}
\item[(b)] for $1 < t < T$:
\begin{equation}
\begin{split}
\log(p(\X_1^i \condit \cdot)) &\propto \log(p(Y_t \condit \mathcal{X}_t, \beta_0)) \\
&-\frac{1}{2 \sigma_{Z_t^i}^2} \norm{\X_t^i - \lambda \bmu_{Z_t^i} - (1 - \lambda) \X_{t-1}^i}_2^2 \\
&-\frac{1}{2 \sigma_{Z_{t+1}^i}^2} \norm{\X_{t+1}^i - \lambda \bmu_{Z_{t+1}^i} - (1 - \lambda) \X_t^i}_2^2,
\end{split}
\end{equation}
\item[(c)] for $t = T$:
\begin{equation}
\begin{split}
\log(p(\X_T^i \condit \cdot)) &\propto \log(p(Y_T \condit \mathcal{X}_T, \beta_0))\\
&-\frac{1}{2 \sigma_{Z_T^i}^2} \norm{\X_T^i - \lambda \bmu_{Z_T^i} - (1 - \lambda) \X_{T-1}^i}_2^2.
\end{split}
\end{equation}
\end{enumerate}
Finally the log of the full conditional for the intercept parameter used in the MH step is
\begin{equation}
\log(p(\beta_0 \condit \cdot)) \propto \log(p(Y_{1:T} \condit \mathcal{X}_{1:T}, \beta_0)) - \frac{1}{2 \sigma_{\beta_0}^2} \norm{\beta_0 - \mu_{\beta_0}}_2^2.
\end{equation}

\subsection{Derivation of the Forward-Backward Algorithm for $Z_{1:T}^i$}\label{subsec:label_sampler}

To jointly sample an actor's group assignments $Z_{1:T}^i$, we use a variant of the forward-backward algorithm for HMMs \citep{rabiner1989}. Recall that the model assumes that each actor's trajectory is an iid AR-HMM conditioned on the transition and emission parameters. To sample the group assignments as a block, we begin by factoring the joint conditional distribution in a forward fashion:
\begin{equation}
p(Z_{1:T}^i \condit \X_{1:T}^i) = p(Z_1^i \condit \X_{1:T}^i) \prod_{t=2}^T p(Z_t^i \condit Z_{t-1}^i, \X_{1:T}^i),
\end{equation}
where we are implicitly conditioning on the model parameters $\bpi_0^1, \Pi_{2:T}, \bmu_{1:L}$, and $\sigma^2_{1:L}$ for clarity. The strategy is then to sample the cluster assignments according to this forward factorization. At the first time step we sample $Z_1^i  \sim p(Z_1^i \condit \X_{1:T}^i)$, and then we draw $Z_t^i \sim p(Z_t^i \condit Z_{t-1}^i, \X_{1:T}^i)$ for $t = 2, \dots, T$. These conditional distributions can be efficiently sampled using backwards message variables. Specifically, we define the backwards message variables recursively as
\begin{equation}
m_{t,t-1}(Z_{t-1}^i) \propto
\begin{cases}
\sum_{Z_{t}^i} p(Z_{t}^{i} \condit Z_{t-1}^i)p(\X_t^i \condit Z_{t}^i, \X_{t-1}^{i})m_{t+1,t}(Z_t^i), & t \leq T, \\
1, & t = T + 1.
\end{cases}
\end{equation}
We can then express the conditional distributions in terms of the backwards message variables as follows:
\begin{equation}
\begin{split}
p(Z_t^i \condit Z_{t-1}^{i}, \X_{1:T}^i) &\propto p(Z_t^i, Z_{t-1}^i \X_{1:T}^i) \\
&= \sum_{Z_{1:(t-2)}^i} \sum_{Z_{(t+1):T}^i} p(Z_{1:T}^i, \X_{1:T}^i) \\
&\propto p(Z_t^i \condit Z_{t-1}^i) p(\X_t^i \condit Z_t^i, \X_{t-1}^i) \sum_{Z_{(t+1):T}^i}\prod_{s=t+1}^T p(Z_{s}^i \condit Z_{s-1}^i) p(\X_s^i \condit Z_s^i, \X_{s-1}^i) \\
&= p(Z_t^i \condit Z_{t-1}^i) p(\X_t^i \condit Z_t^i, \X_{t-1}^i)\sum_{Z_{t+1}^i} p(Z_{t+1}^i \condit Z_t^i) p(\X_{t+1}^i \condit Z_{t+1}^i, \X_t^i) m_{t+2, t+1}(Z_{t+1}^i) \\
&= p(Z_t^i \condit Z_{t-1}^i) p(\X_t^i \condit Z_t^i, \X_{t-1}^i) m_{t+1, t}(Z_{t}^i).
\end{split}
\end{equation}
The full label sampling algorithm is detailed in Algorithm \ref{alg:label_sampler}.



%
%

\subsection{Hyperparameter Samplers}\label{subsec:hyper_appendix}

In this section, we present the samplers for the hyperparamters of the HDP-LPCM. The full conditionals for the prior variance of the group means $\tau^2$ and the prior scale of the group variances $b$ are conjugate given our choice of hyperpriors. These full conditionals are displayed in Equations (\ref{eq:tau_posterior}) and (\ref{eq:b_posterior}) respectively. The remaining hyperparameters are the concentration parameters of the HDP. To sample these hyperparameters, we use the auxiliary variable samplers developed in \citet{teh2006} and \citet{fox2011a}, which we slightly modified to account for the resampling of transition distributions. These samplers utilize the auxiliary variable sampler developed by \citet{escobar1995}. For completeness, we include this sampler in Algorithm \ref{alg:escobar_west_sampler}. Our full hyperparameter sampler is detailed in Algorithm \ref{alg:hyperparameters}.

\begin{sampler}[h]
The auxiliary variable sampler for the concentration parameter $\gamma$ in a $\DP(\gamma, H)$ presented in \citet{escobar1995}. Note that a $\Gamma(a, b)$ hyperprior is placed on $\gamma$.

\begin{enumerate}
\item Sample $\eta \sim \Beta(\gamma + 1, N)$.

\item Define $\tilde{\pi}_1 = a + K - 1$, $\tilde{\pi}_2 = N (b - \log(\eta))$, then define $\pi = \tilde{\pi}_1 / (\tilde{\pi}_1 + \tilde{\pi}_2)$.
\item Sample the concentration parameter from a mixture of Gamma distributions
\begin{equation*}
\gamma \sim \pi \Gamma(a + K, b - \log(\eta)) + (1 - \pi) \Gamma(a + K - 1, b - \log(\eta)).
\end{equation*}
\end{enumerate}

\caption{Auxiliary variable sampler used to sample concentration parameters developed by \citet{escobar1995}.}
\label{alg:escobar_west_sampler}

\end{sampler}

\begin{sampler}[htbp]
Note that we set the following priors on the HDP parameters:
\begin{equation*}
\begin{split}
\gamma &\sim \Gamma(a_{\gamma}, b_{\gamma}), \\
\alpha_0 &\sim \Gamma(a_{\alpha_0}, b_{\alpha_0}), \\
\alpha + \kappa &\sim \Gamma(a_{\alpha + \kappa}, b_{\alpha + \kappa}), \\
\rho &\sim \Beta(a_{\rho}, b_{\rho}).
\end{split}
\end{equation*}

Given a previous set of hyperparameters $(\tau^2)^{(\ell - 1)}$, $b^{(\ell - 1)}$, $\gamma^{(\ell -1)}$, $\alpha_0^{(\ell - 1)}$, $\alpha^{(\ell - 1)}$, and $\kappa^{(\ell - 1)}$, update the current parameters as follows:

\begin{enumerate}
\item Initialize current parameters to the values of the $(\ell - 1)$th iteration.

\item Update prior variance:
\begin{equation*}
\tau^2 \sim \InvGamma((a_{\tau} + L) / 2, (b_{\tau} + \sum_{g=1}^L \norm{\bmu_g}_2^2) / 2).
\end{equation*}

\item Update prior scale:
\begin{equation*}
b \sim \Gamma((c + L \, a )/2 , 2 (d + \sum_{g=1}^L \sigma_g^{-2})^{-1}).
\end{equation*}

\item Update $\gamma$ via Algorithm \ref{alg:escobar_west_sampler} with $K = \sum_{k=1}^L \ind{\bar{m}_{\cdot k \cdot} > 0}$, $N = \bar{m}_{\cdot \cdot \cdot}$, $a = a_{\gamma}$, and $b = b_{\gamma}$.

\item Update $\alpha_0$ via Algorithm \ref{alg:escobar_west_sampler} with $K = m_{0\cdot1}$,  $N = n$, $a = a_{\alpha + \kappa}$, and $b = b_{\alpha + \kappa}$.

\item Update $\alpha + \kappa$:

For $g \in \set{1, \dots, L}$ and $t \in \set{2, \dots, T}$ sample
\begin{equation*}
\begin{split}
r_{gt} &\sim \Beta(\alpha + \kappa + 1, n_{g\cdot t}), \\
s_{gt} &\sim \Bern(\frac{n_{g\cdot t}}{n_{g\cdot t} + \alpha + \kappa}).
\end{split}
\end{equation*}
Update concentration parameter:
\begin{equation*}
\alpha + \kappa \sim \Gamma(a_{\alpha + \kappa} + m_{\cdot \cdot \cdot} - s_{\cdot \cdot}, b_{\alpha + \kappa} - \sum_{g=1}^L \sum_{t=2}^T \log(r_{gt})).
\end{equation*}

\item Update $\rho$:
\begin{equation*}
\rho \sim \Beta(w_{\cdot \cdot \cdot} + a_{\rho}, \sum_{s=2}^T m_{\cdot \cdot s} - w_{\cdot \cdot \cdot} + b_{\rho}),
\end{equation*}
where we exclude the considered dishes at $s = 1$ because the initial restaurant does not include a stickiness parameter.

\end{enumerate}
\caption{Samplers for the hyperparameters in the HDP-LPCM. In all applications, we set $a_{\alpha + \kappa} = 5$, $b_{\alpha + \kappa} = 0.1$, $a_{\rho} = 8$, $b_{\rho} = 2$, $a_{\gamma} = 1$, $b_{\gamma} = 0.1$, $a_{\alpha_0} = 1$, and $b_{\alpha_0} = 1$.}
\label{alg:hyperparameters}
\end{sampler}

\clearpage

\subsection{Sampson's Monastery}\label{subsec:sampson}

In this section, we apply the HDP-LPCM to dynamic networks of social relations between 18 monks in an isolated New England monastery \citep{sampson}. The networks consists of `liking' relations among the monks, as measured by Sampson for his doctoral dissertation. Specifically, as a resident at the monastery, Sampson asked each monk to name up to three other monks that he liked most on three different occasions during his study. Based on these measurements, Sampson divided the monks into three tight factions: the Loyal Opposition (7 members), the Outcasts (4 members), and the Young Turks (7 members). During his stay, the existence of three factions was confirmed when a `crisis in the cloister' occurred, resulting in the expulsion of the leaders of the Young Turks (John Bosco and Gregory) and then the voluntary departure of several others. For this reason, Sampson's labels are often used as ground truth in the network literature.

We analyze the undirected binary version of Sampson's networks. In this case, an edge between monks $i$ and $j$ exists at time $t$ if monk $i$ nominated monk $j$ (or vice versa) as a friend during time $t$. The result is $T = 3$ undirected networks with $n = 18$ actors. Since this dataset is commonly analyzed using latent position and latent cluster models, we will compare our results to findings of previous analyses to critique our model's performance. In addition, we extend past analyses by providing a measure of stability for the network's group structure. In particular, we aim to answer the question of whether the three factions were present over all observational periods or split-off during a particular time point.

\begin{figure}[hp]
\centering
\includegraphics[height=0.6\textheight]{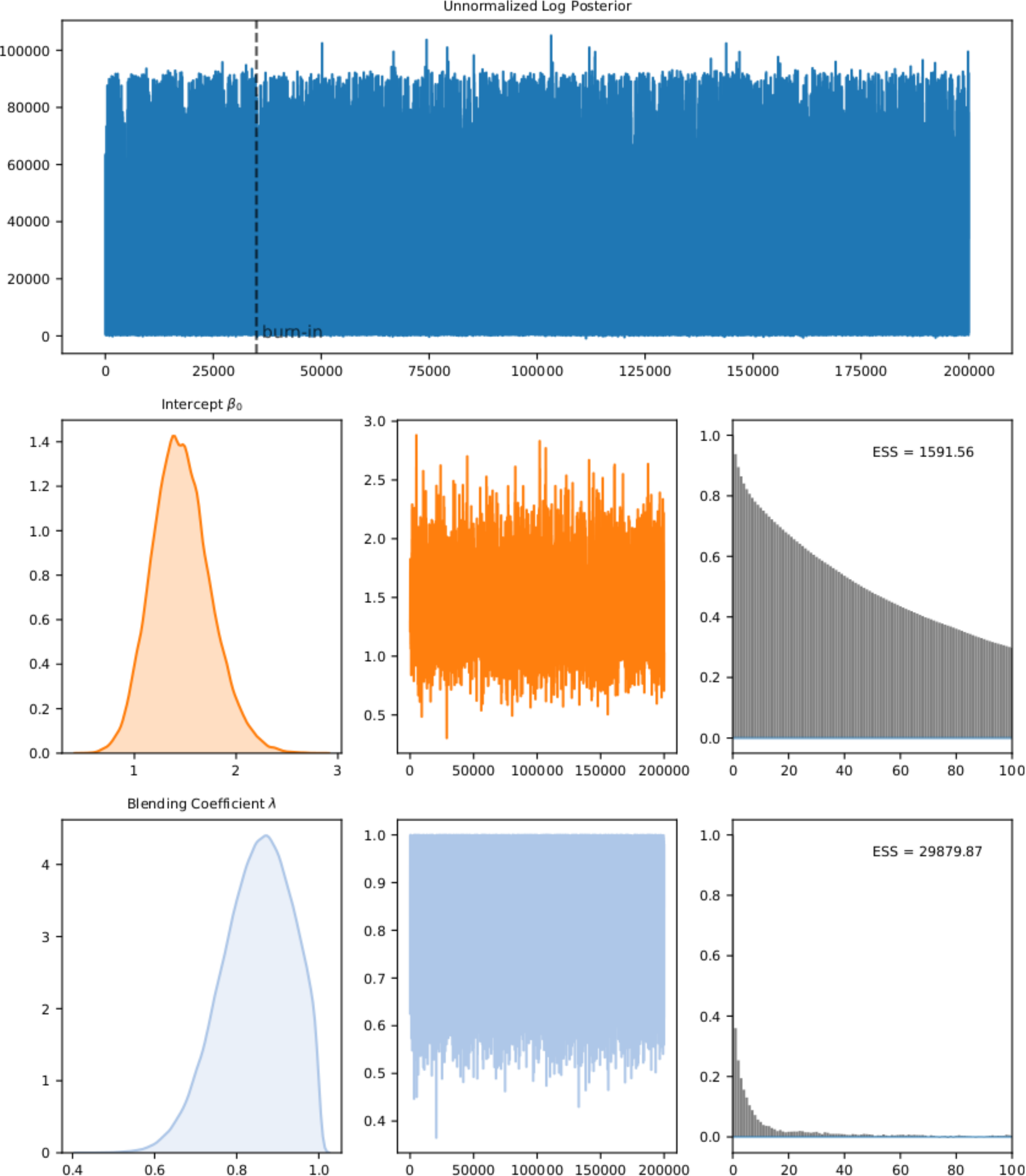}
\caption{Various diagnostic plots for the MCMC algorithm used to analyze Sampson's monastery network. Trace plot of the unnormalized posterior value of each iteration of the MCMC chain (first row). Kernel density estimate of the marginal posterior, trace plot, and ACF plot for $\beta_0$ (second row), and $\lambda$ (third row). The effective sample size (ESS) of the $\beta_0$ and $\lambda$ chains are displayed in the upper right corners of the ACF plots.}
\label{fig:sampson_traces}
\end{figure}

We fit the HDP-LPCM using 15,000 iterations for tuning, 20,000 iterations for burn-in, which left a final chain of length 165,000 iterations for inference. The trace plots of the unnormalized log-posterior, the intercept $\beta_{0}$ and the blending coefficient $\lambda$ are displayed in Figure \ref{fig:sampson_traces}. We also provide ACF plots and a kernel density estimate of the marginal posterior distributions. Visual inspection of the trace plots indicates that the algorithm has converged. 

\begin{figure}[htbp]
    \begin{subfigure}[b]{0.32\textwidth}
    \includegraphics[width=\textwidth]{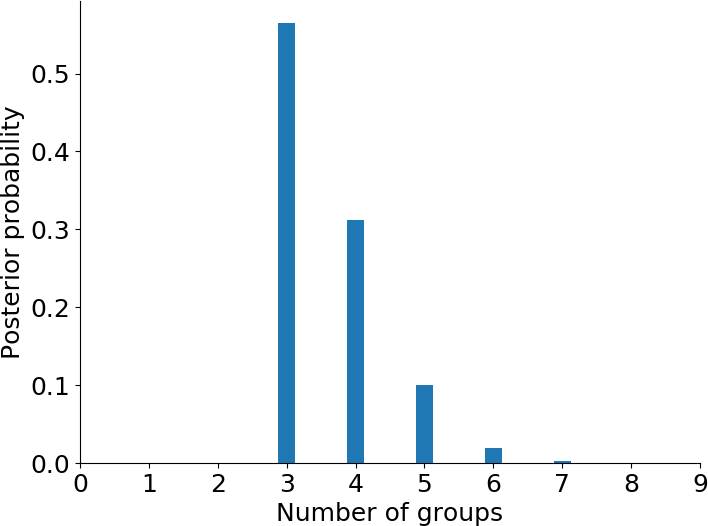}
    \caption{$t = 1$}
    \label{fig:sampson_groupsa}
    \end{subfigure}
    \hfill
    \begin{subfigure}[b]{0.32\textwidth}
    \includegraphics[width=\textwidth]{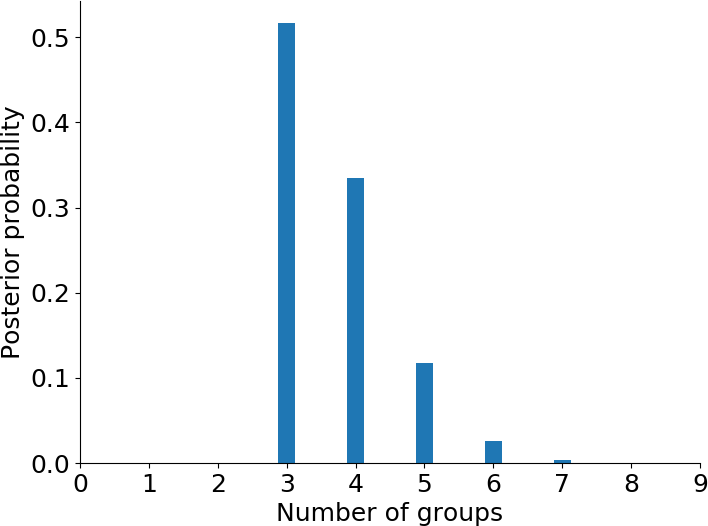}
    \caption{$t = 2$}
    \label{fig:sampson_groupsb}
    \end{subfigure}
    \hfill
    \begin{subfigure}[b]{0.32\textwidth}
    \includegraphics[width=\textwidth]{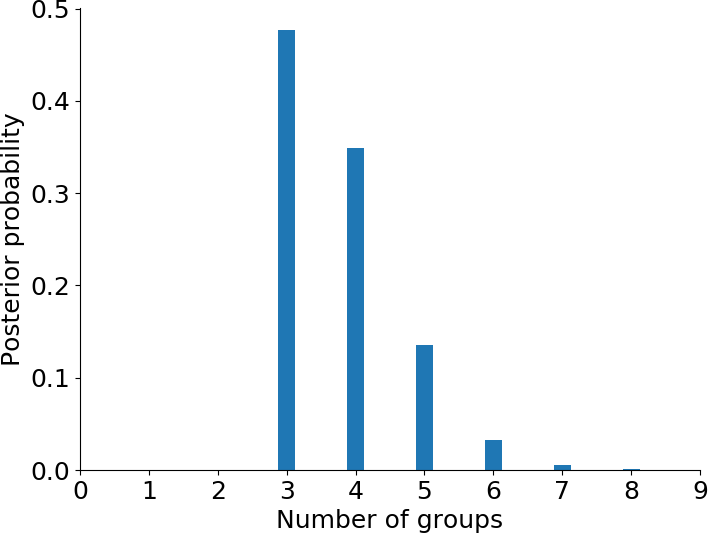}
    \caption{$t = 3$}
    \label{fig:sampson_groupsc}
    \end{subfigure}
    \caption{(a) - (c) Posterior probabilities over the number of groups for Sampson's monastery network at times $t = 1, 2, 3$. For each time step the posterior places the most mass on $G_t = 3$ groups. This indicates that the monastery has a static group structure.}
    \label{fig:sampson_groups}
\end{figure}

Our analysis corroborates many of the previous findings about the group structure of Sampson's monastery. The posterior mean of the blending coefficient $\lambda$ is 0.869, which confirms Sampson's assertion that a strong group effect drives the dynamics of the monk's social network. Furthermore, previous static and dynamic analysis widely agree that a 3 or 4 component model adequately describes the network \citep{handcock2007, krivitsky2009, ryan2017}. The posterior distribution over the number of groups at each time step $G_t$ is displayed in Figures \ref{fig:sampson_groupsa} - \ref{fig:sampson_groupsc}. These distributions confirm that 3 groups is the most probable number of groups at each point in time.

\begin{figure}[htbp]
\centering
\includegraphics[height=0.25\textheight]{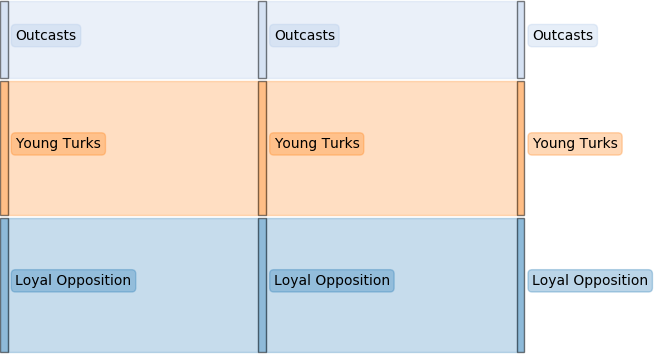}
\caption{Alluvial diagram showing the dynamics of the group memberships estimated by our model on Sampson's monastery network. Each line represents the flux of nodes going from one group to the next at time $t$ to $t+1$. The thickness of the lines is proportional to the number of nodes and the total height represents all nodes. In the case of Sampson's monastery, the group memberships of each monk is static over all time periods.}
\label{fig:sampson_alluvial}
\end{figure}

\begin{figure}[hp]
    \centering
    \begin{subfigure}[b]{0.45\textwidth}
        \centering
        \includegraphics[width=\textwidth]{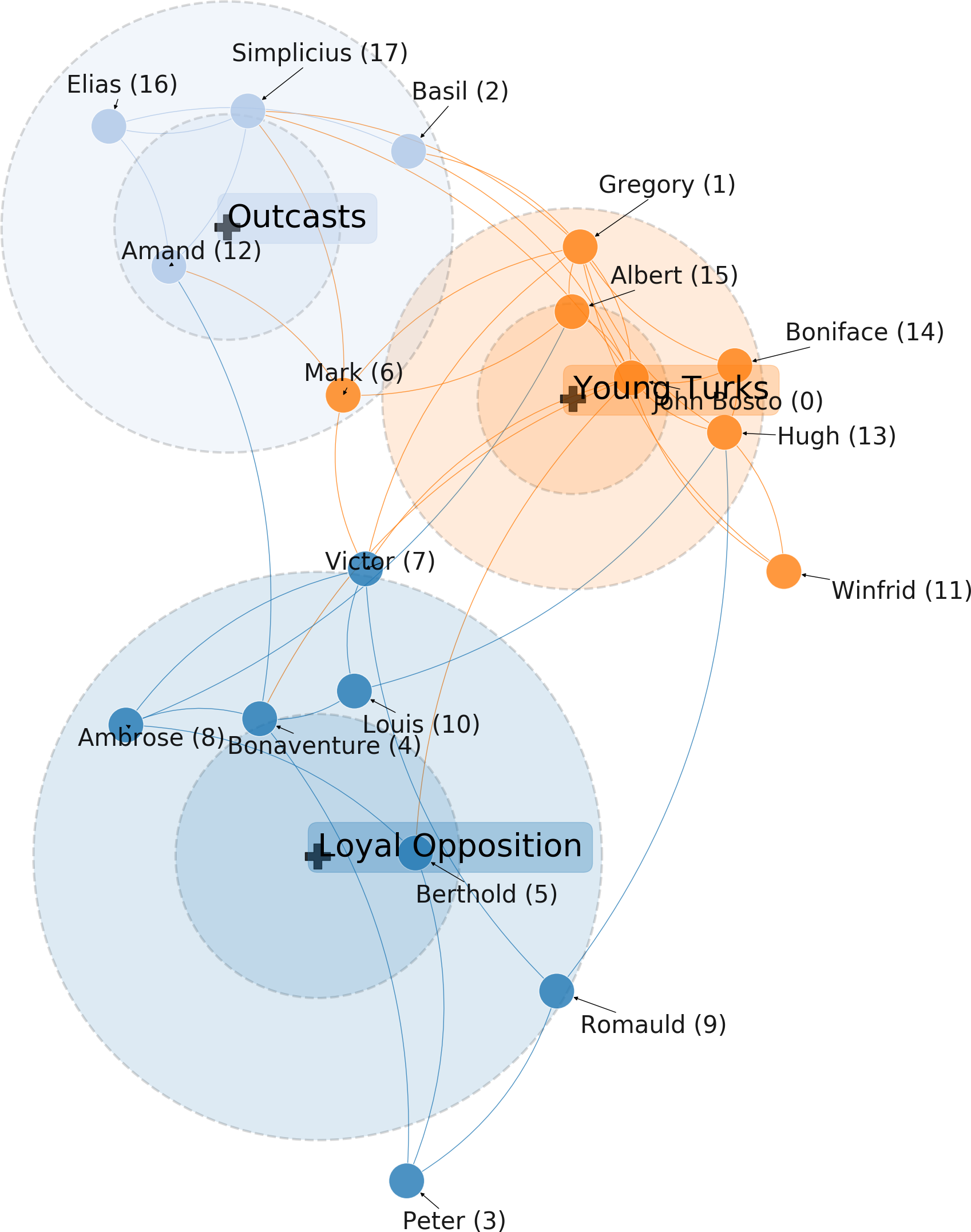}
        \caption*{$t = 1$}
    \end{subfigure}
    \hfill
    \begin{subfigure}[b]{0.45\textwidth}
        \centering
        \includegraphics[width=\textwidth]{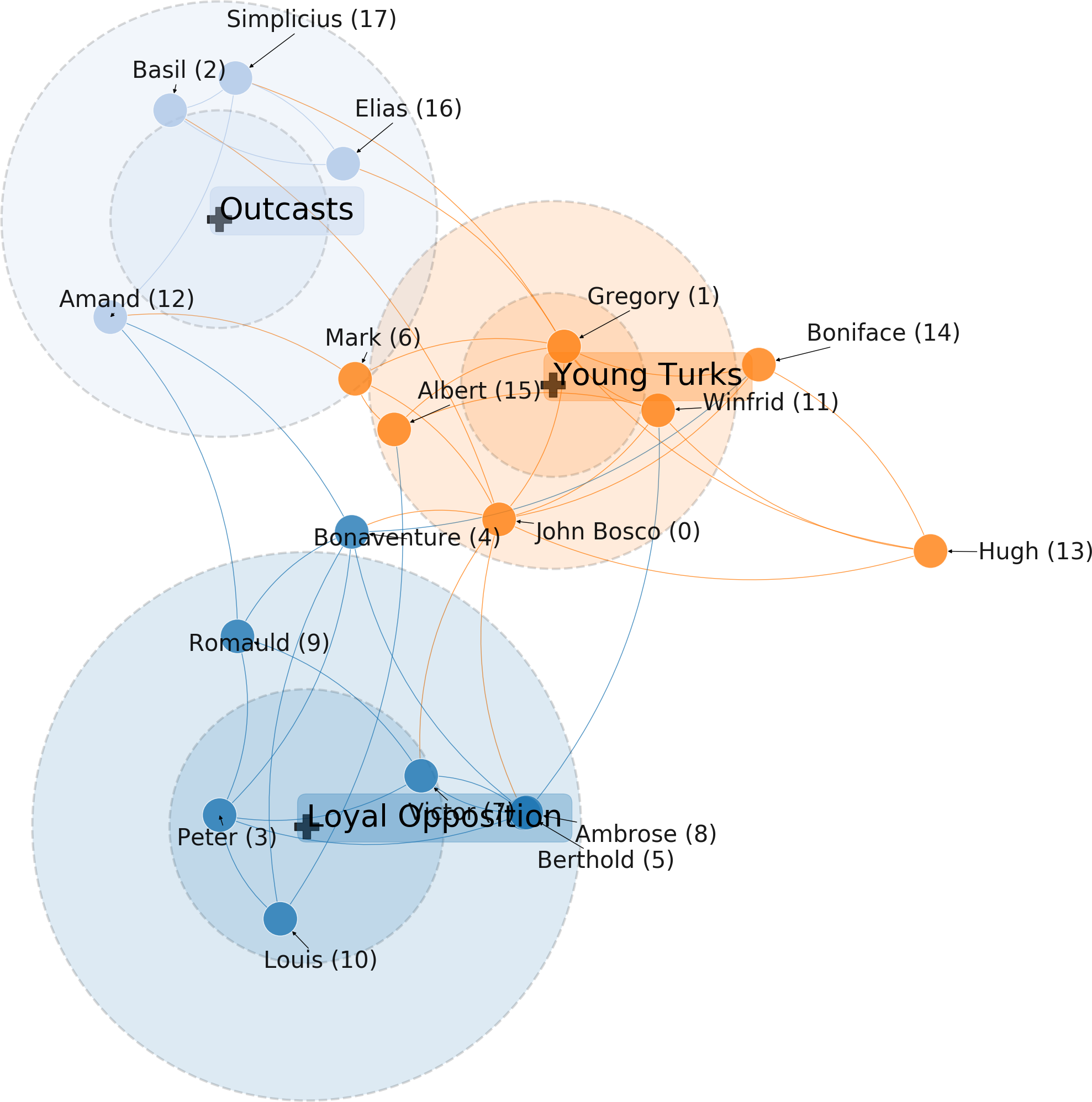}
        \caption*{$t = 2$}
    \end{subfigure}
    \begin{subfigure}[b]{0.45\textwidth}
        \centering
        \includegraphics[width=\textwidth]{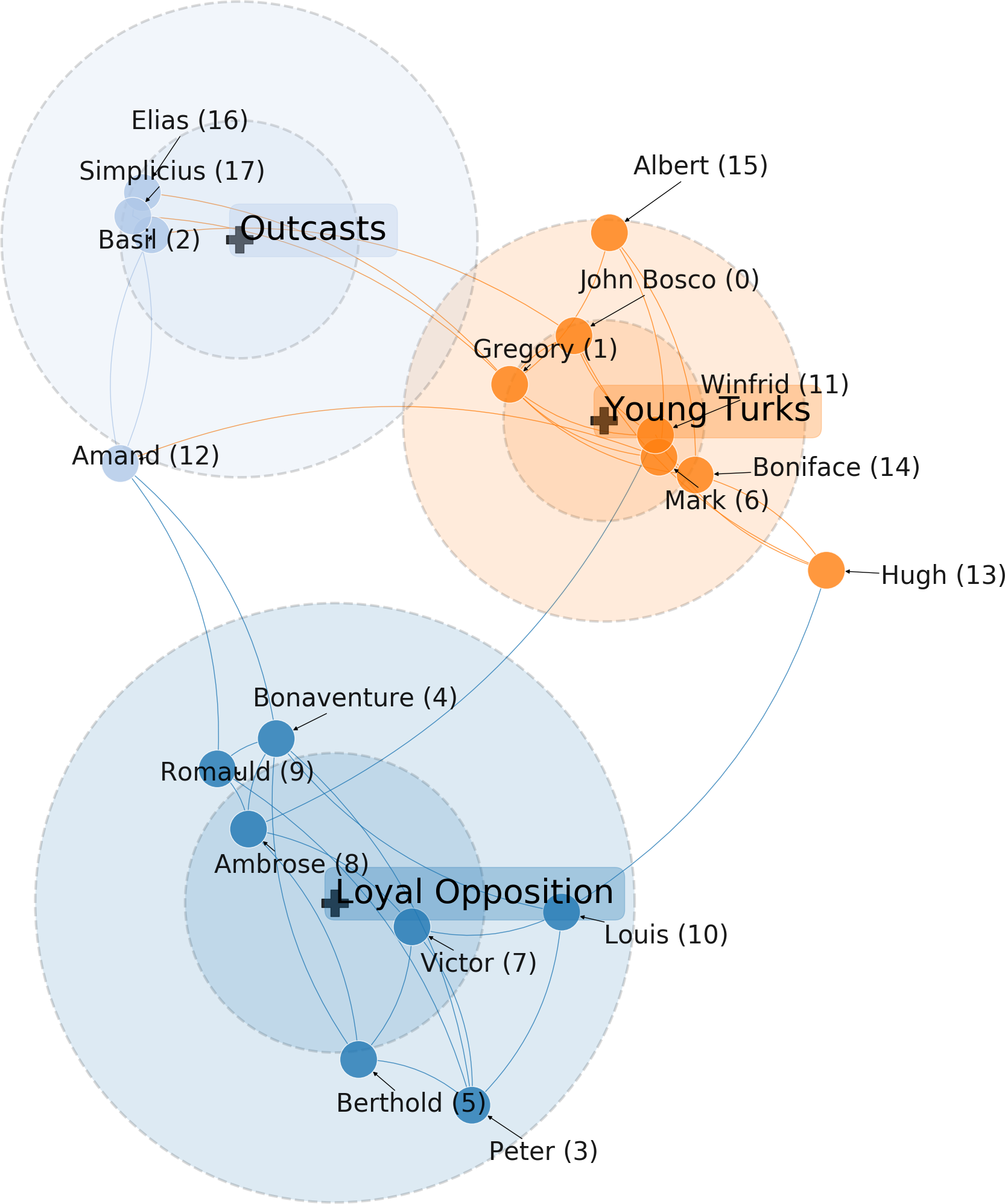}
        \caption*{$t = 3$}
    \end{subfigure}
    \caption{Estimated latent space for Sampson's monastery network. The three groups inferred by the model are labeled according to the corresponding faction in Sampson's ground truth labels (Loyal Opposition, Young Turks, Outcasts). The group means $\bmu_g$ are denoted by a $+$, and the group shapes $\sigma_g$ are displayed as two-standard deviation ellipses. The names of each monk and node number are annotated. The undirected edges are also displayed.}
    \label{fig:sampson_latent_space}
\end{figure}

A distinctive advantage of our approach compared to other latent space models is the ability to quantify the stability of the factions within the monastery. To do this, we follow the decision theoretic procedure described in Section \ref{subsec:model_selection} to select a representative partition that summarizes the posterior distribution. This partition's associated alluvial diagram (Figure \ref{fig:sampson_alluvial}) and latent space (Figure \ref{fig:sampson_latent_space}) both indicate that there is a stable three group structure during all time points. Furthermore, we observe that the monk's group memberships do not change over time. As further confirmation of our method, we note that the HDP-LPCM assigns all monks to the same communities as Sampson's ground truth labels. Overall, our method is not only consistent with previous latent space analysis of Sampson's monastery network, but also expands the literature through its indication that the group structure is static over the observed time period.

\clearpage

\subsection{Additional Figures}\label{subsec:additional_figures}

In this section, we include the remaining figures for the two datasets analyzed in the real data applications of the main text (Section \ref{sec:real_data}). This includes the trace plots for the MCMC algorithm as well as additional latent space visualizations.

\begin{figure}[hp]
\centering
\includegraphics[height=0.6\textheight]{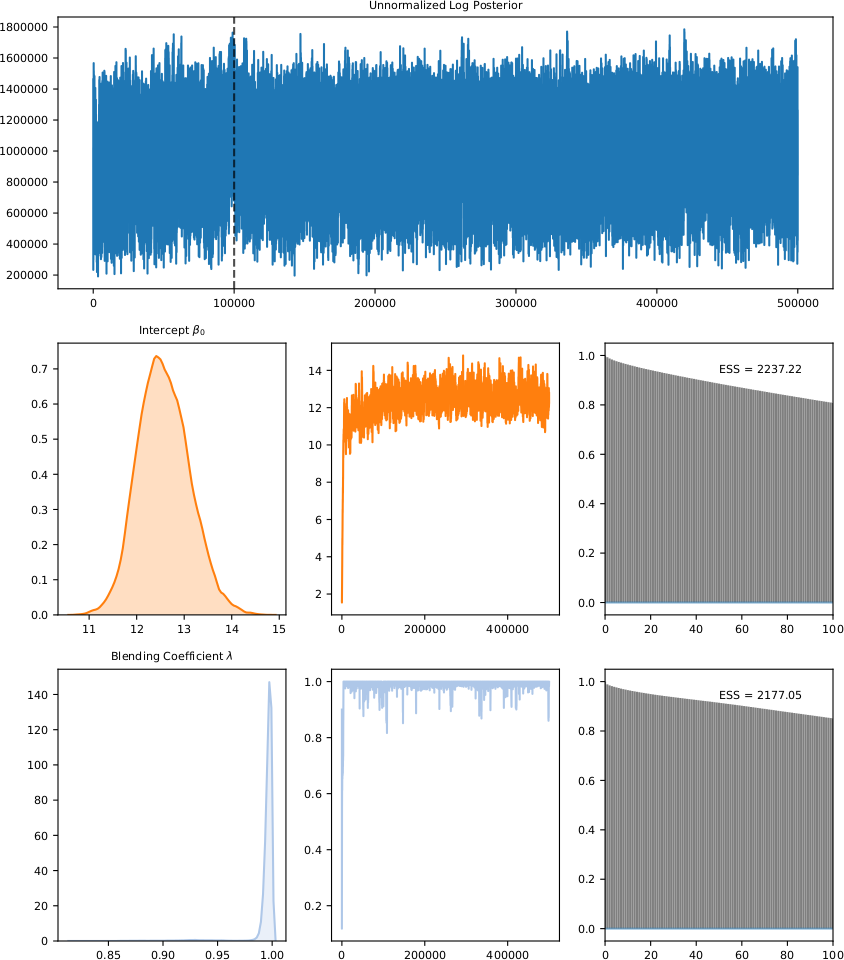}
\caption{Various diagnostic plots for the MCMC algorithm used to analyze international military alliances network. Trace plot of the unnormalized posterior value of each iteration of the MCMC chain (first row). Kernel density estimate of the marginal posterior, trace plot, and ACF plot for $\beta_0$ (second row), and $\lambda$ (third row). The effective sample size (ESS) of the $\beta_0$ and $\lambda$ chains are displayed in the upper right corners of the ACF plots.}
\label{fig:alliances_traces}
\end{figure}

\begin{figure}[htbp]
    \includegraphics[width=\textwidth]{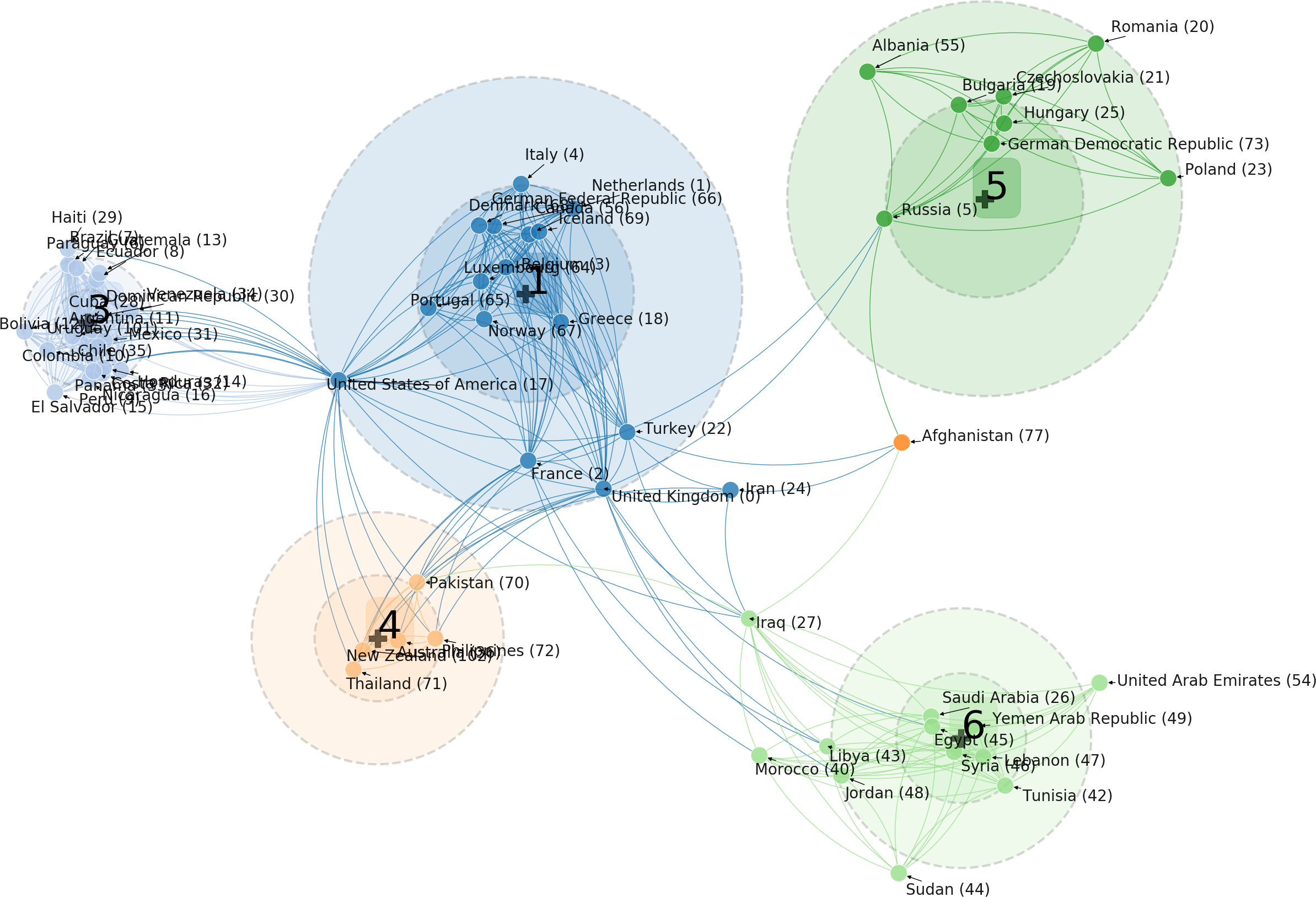}
    \caption{Latent space of international military alliances for the years 1955 - 1959. The group means $\bmu_g$ are denoted by a $+$, and the group shapes $\sigma_g$ are displayed as two-standard deviation ellipses. The names of each nation and node number are annotated. The undirected edges are also displayed. For clarity, all unconnected nodes and group 2's two-standard deviation ellipses are removed.}
    \label{fig:alliances_t1}
\end{figure}

\begin{figure}[htbp]
    \includegraphics[width=\textwidth]{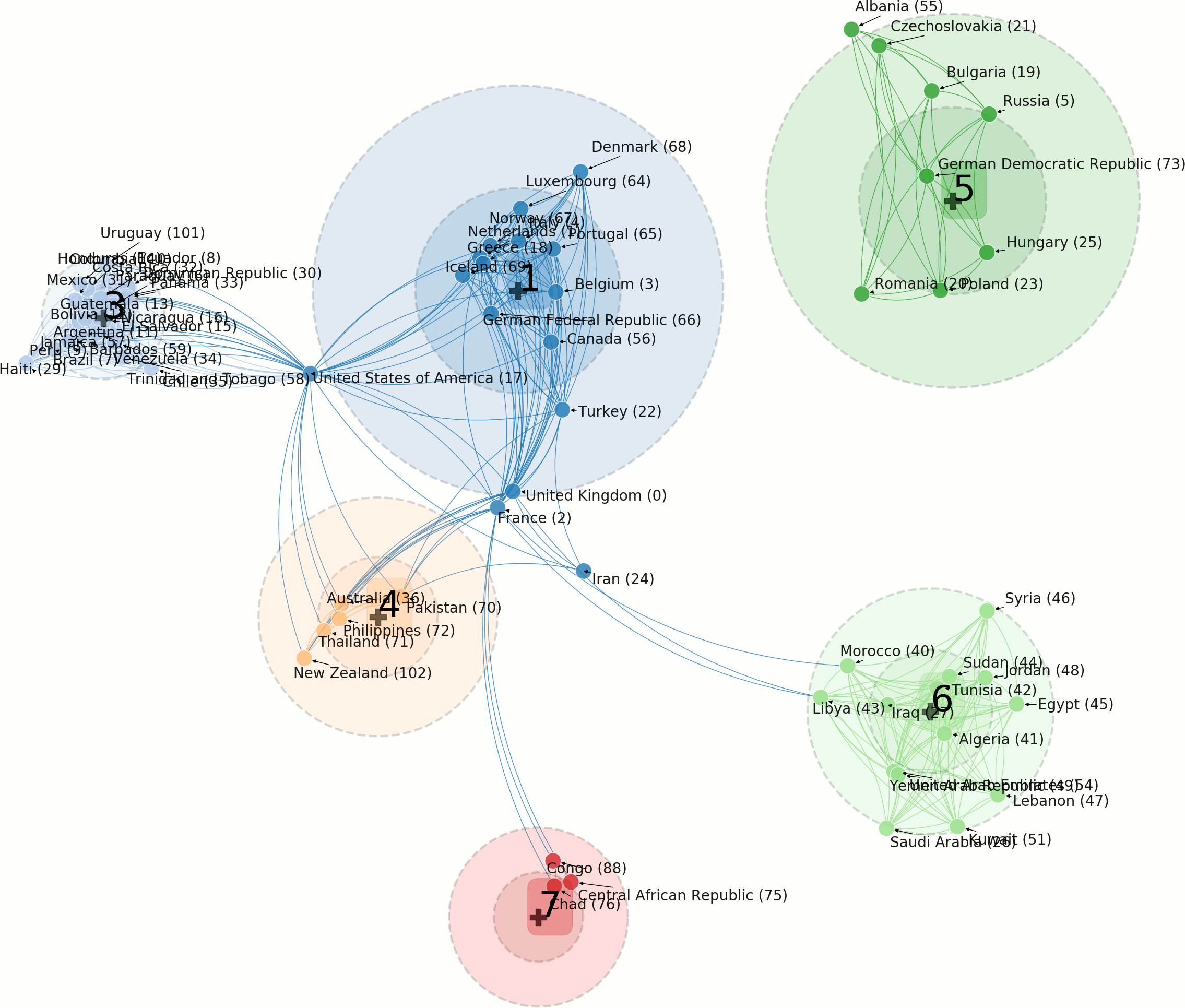}
    \caption{Latent space of international military alliances for the years 1965 - 1969. The group means $\bmu_g$ are denoted by a $+$, and the group shapes $\sigma_g$ are displayed as two-standard deviation ellipses. The names of each nation and node number are annotated. The undirected edges are also displayed. For clarity, all unconnected nodes and group 2's two-standard deviation ellipses are removed.}
    \label{fig:alliances_t3}
\end{figure}

\begin{figure}[htbp]
    \includegraphics[width=\textwidth]{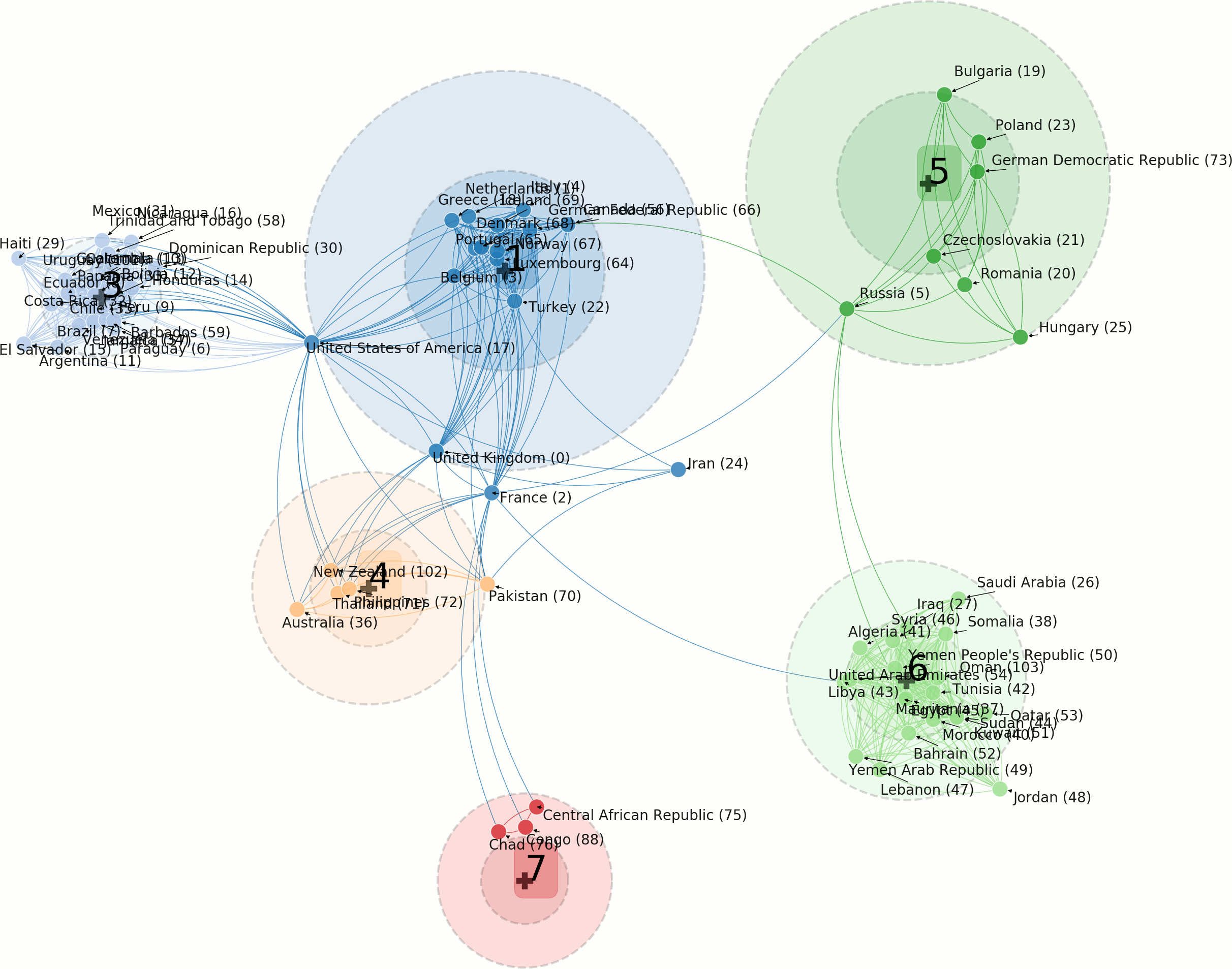}
    \caption{Latent space of international military alliances for the years 1970 - 1974. The group means $\bmu_g$ are denoted by a $+$, and the group shapes $\sigma_g$ are displayed as two-standard deviation ellipses. The names of each nation and node number are annotated. The undirected edges are also displayed. For clarity, all unconnected nodes and group 2's two-standard deviation ellipses are removed.}
    \label{fig:alliances_t4}
\end{figure}

\begin{figure}[htbp]
    \includegraphics[width=\textwidth]{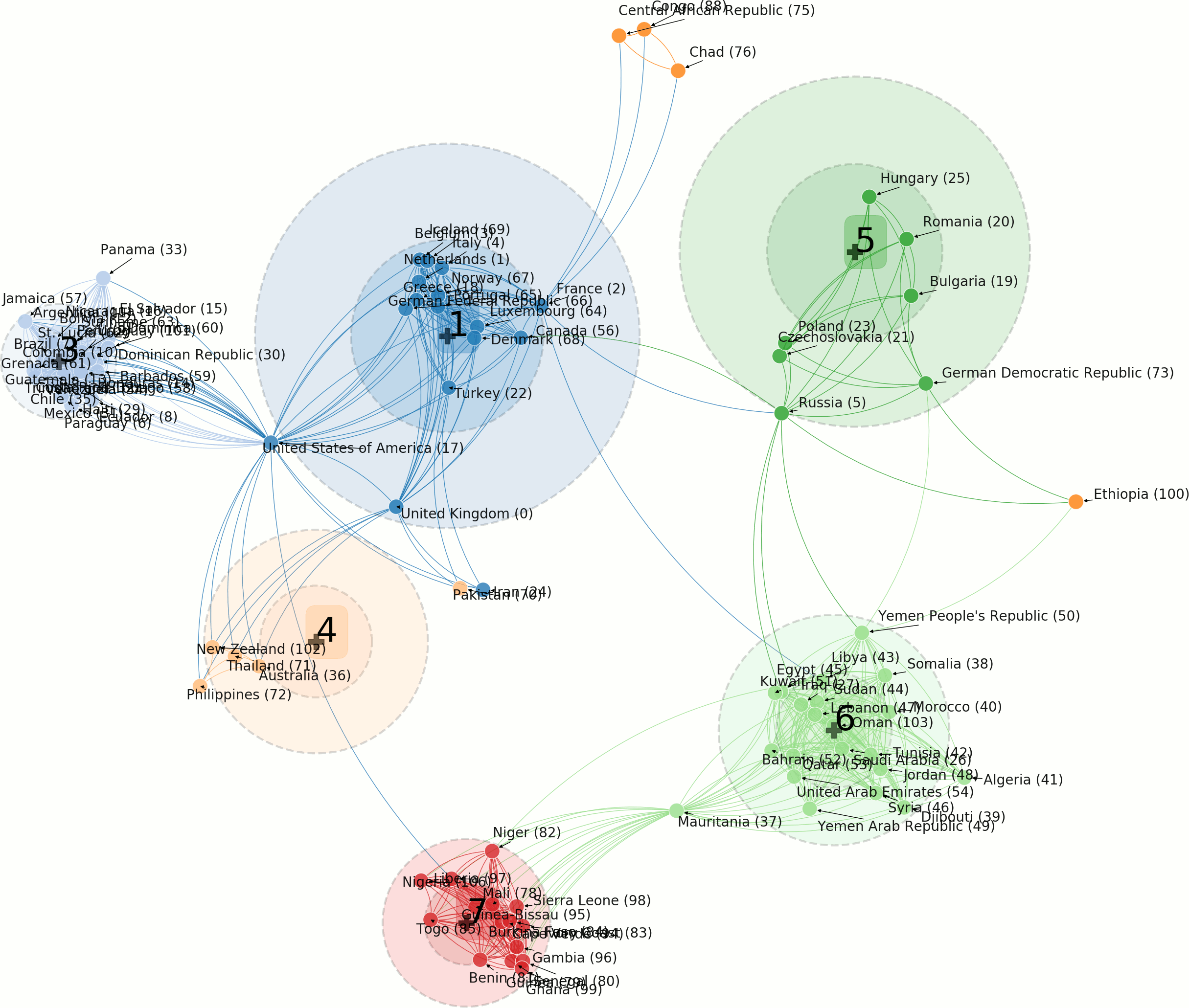}
    \caption{Latent space of international military alliances for the years 1975 - 1979. The group means $\bmu_g$ are denoted by a $+$, and the group shapes $\sigma_g$ are displayed as two-standard deviation ellipses. The names of each nation and node number are annotated. The undirected edges are also displayed. For clarity, all unconnected nodes and group 2's two-standard deviation ellipses are removed.}
    \label{fig:alliances_t5}
\end{figure}

The trace plots of the unnormalized log-posterior, the intercept $\beta_{0}$ and the blending coefficient $\lambda$ for the HDP-LPCM fit to the international military alliances network are displayed in Figure \ref{fig:alliances_traces}. The latent spaces for 1955-1959, 1965-1969, 1970-1974, and 1975-1979 are displayed in Figures \ref{fig:alliances_t1}, \ref{fig:alliances_t3}, \ref{fig:alliances_t4}, \ref{fig:alliances_t5} respectively.

\begin{figure}[hp]
\centering
\includegraphics[height=0.6\textheight]{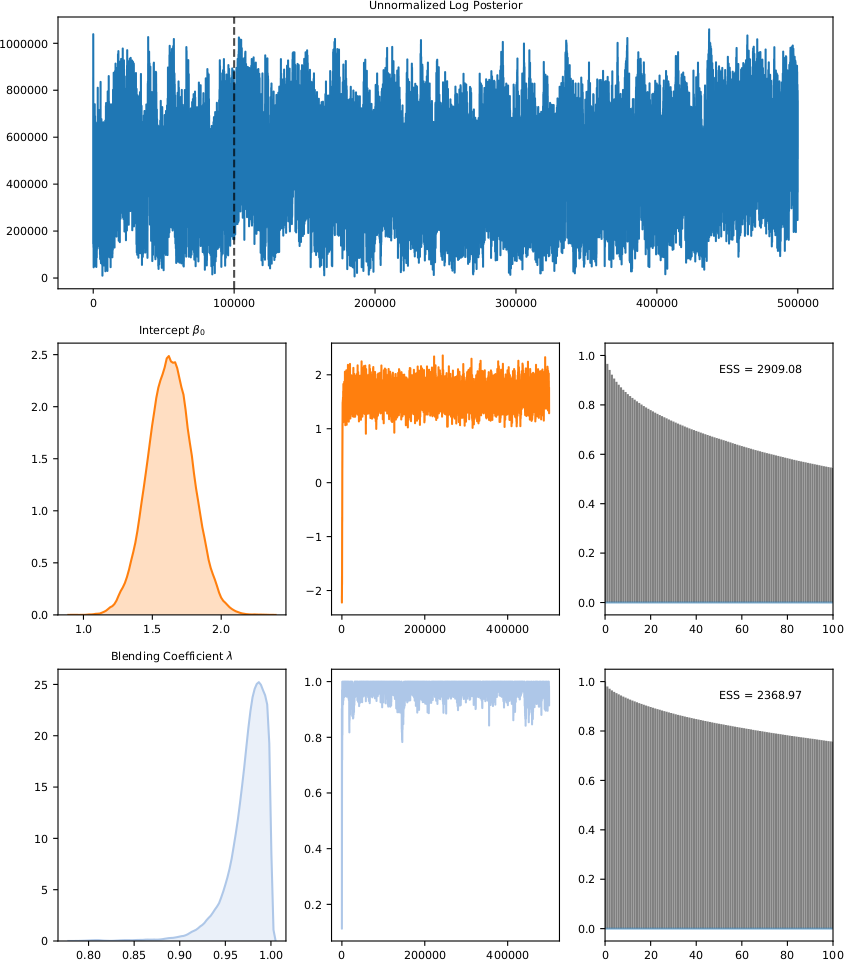}
\caption{Various diagnostic plots for the MCMC algorithm used to analyze Game of Thrones network. Trace plot of the unnormalized posterior value of each iteration of the MCMC chain (first row). Kernel density estimate of the marginal posterior, trace plot, and ACF plot for $\beta_0$ (second row), and $\lambda$ (third row). The effective sample size (ESS) of the $\beta_0$ and $\lambda$ chains are displayed in the upper right corners of the ACF plots.}
\label{fig:got_traces}
\end{figure}

\begin{figure}[htbp]
    \centering
    \includegraphics[width=\textwidth]{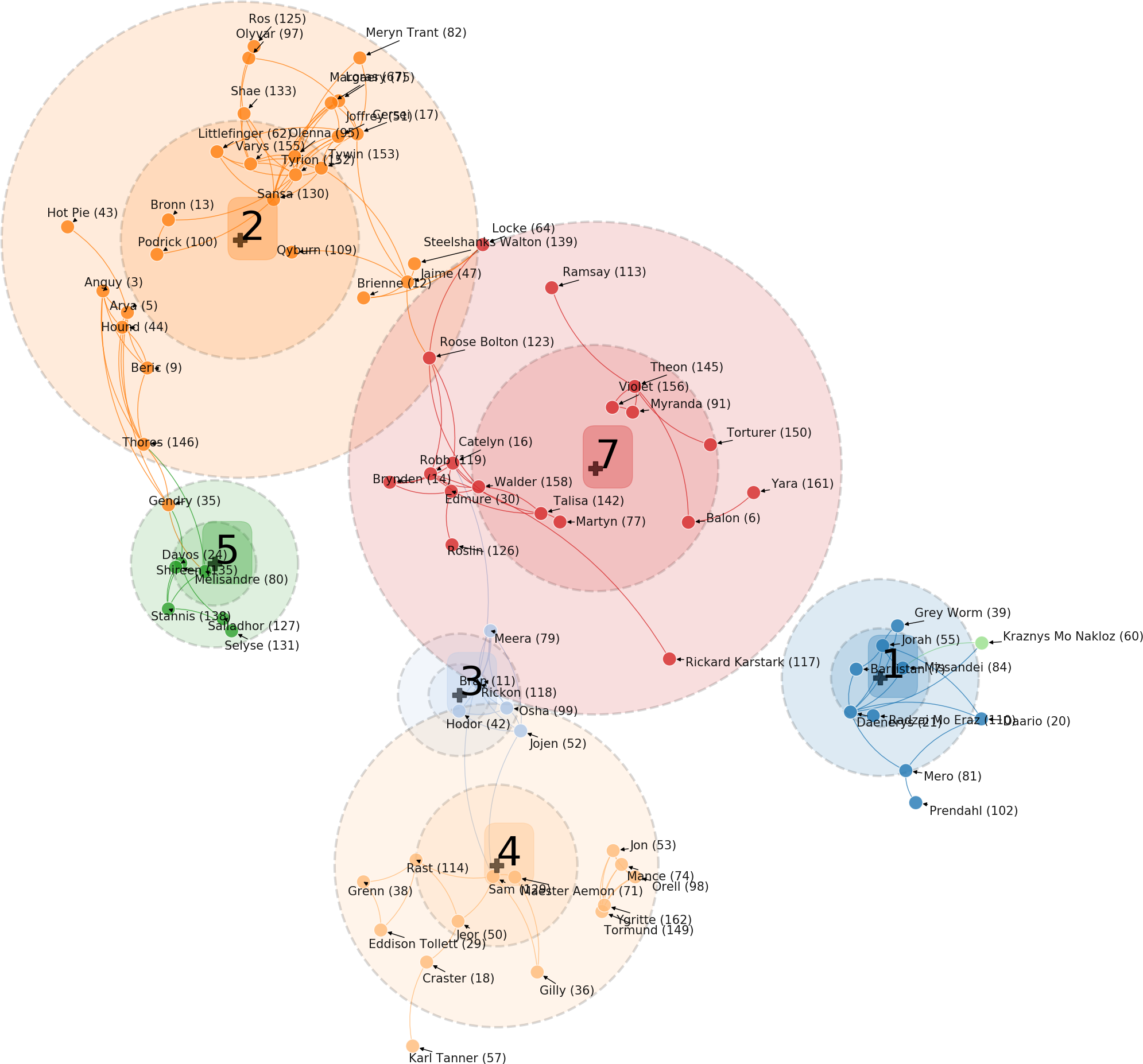}
    \caption{Latent space of the Game of Thrones character interaction network for season 3. The group means $\bmu_g$ are denoted by a $+$, and the group shapes $\sigma_g$ are displayed as two-standard deviation ellipses. The names of each character and node number are annotated. The undirected edges are also displayed. For clarity, all unconnected nodes and group 6's two-standard deviation ellipses are removed.}
    \label{fig:got_season3}
\end{figure}

\begin{figure}[htbp]
    \centering
    \includegraphics[width=\textwidth]{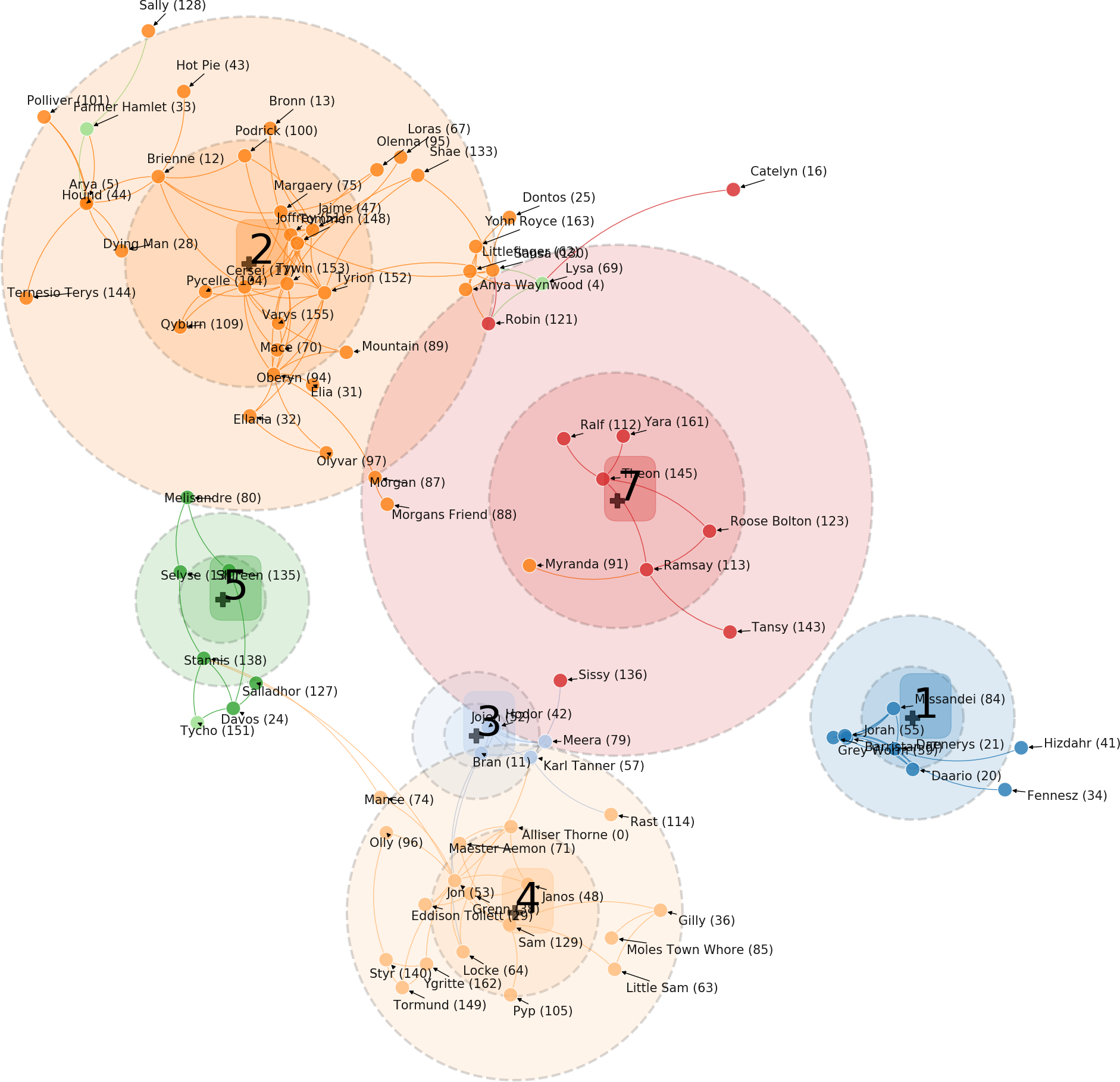}
    \caption{Latent space of the Game of Thrones character interaction network for season 4. The group means $\bmu_g$ are denoted by a $+$, and the group shapes $\sigma_g$ are displayed as two-standard deviation ellipses. The names of each character and node number are annotated. The undirected edges are also displayed. For clarity, all unconnected nodes and group 6's two-standard deviation ellipses are removed.}
    \label{fig:got_season4}
\end{figure}

The trace plots of the unnormalized log-posterior, the intercept $\beta_{0}$ and the blending coefficient $\lambda$ for the HDP-LPCM fit to the Game of Thrones networks are displayed in Figure \ref{fig:got_traces}. The latent spaces for seasons 3 and 4 are displayed in Figures \ref{fig:got_season3} and \ref{fig:got_season4} respectively.

\end{document}